\documentclass[10pt]{wlscirep}
\usepackage[utf8]{inputenc}
\usepackage{graphicx}
\usepackage[T1]{fontenc}
\title{Socioeconomic correlations of urban patterns inferred from aerial images: interpreting activation maps of Convolutional Neural Networks}

\author[1,*]{Jacob Levy Abitbol}
\author[2,1,*]{M\'arton Karsai}
\affil[1]{ \quad Univ Lyon, Inria, CNRS, ENS de Lyon, Universit\'e Claude Bernard Lyon 1, LIP UMR 5668, France}
\affil[2]{ \quad Department of Network and Data Science, Central European University, 1051, Budapest, Hungary}

\affil[*]{Correspondence should be addressed to Jacob Levy Abitbol (jacobolevyabitbol64@gmail.com) and M\'arton Karsai (karsaim@ceu.edu)}

\keywords{socioeconomic status inference, urban patterns, remote sensing, interpretability of convolutional neural networks}

\begin{abstract}
Urbanisation is a great challenge for modern societies, promising better access to economic opportunities while widening socioeconomic inequalities. Accurately tracking how this process unfolds has been challenging for traditional data collection methods, while remote sensing information offers an alternative to gather a more complete view on these societal changes. By feeding a neural network with satellite images one may recover the socioeconomic information associated to that area, however these models lack to explain how visual features contained in a sample, trigger a given prediction. Here we close this gap by predicting socioeconomic status across France from aerial images and interpreting class activation mappings in terms of urban topology. We show that the model disregards the spatial correlations existing between urban class and socioeconomic status to derive its predictions. These results pave the way to build interpretable models, which may help to better track and understand urbanisation and its consequences.
\end{abstract}
\begin{document}

\flushbottom
\maketitle
%
%
\thispagestyle{empty}

\section{Introduction}

Cities have become the economic bedrock of modern nations; in little more than a century they have gone from concentrating 13\% to an estimated 55\% of the world population with 600 of them currently accounting for around half of the global economic output (\$34 trillion of GDP)~\cite{united20182018}. This transition will likely be accelerated in the coming years as an estimated three billion people will move into cities by 2030. This increased urbanisation is in turn a key driver of development as cities provide the national platform for shared prosperity. Indeed, the concentration of people in cities generates agglomeration economies, where the sheer population density facilitates the moving of goods, people, and ideas by removing the physical spaces between people and firms and thus increasing the returns of urban proximity~\cite{glaeser}.

Nevertheless, while urbanisation can entail economic dynamism and social development, it can also create enormous social challenges. The management of natural hazards and pollution, the exclusion of the poor from the city’s socioeconomic fabric and the subsequent surge of social and economic inequalities have become some of the pressing issues that modern metropolises  need to address. This last issue is especially acute in some of the cities with the greatest concentration of wealth. For instance, Gourevitch \emph{et al.} reported a 24 years difference in the average lifespan of citizens living in neighboring census tracks in San Francisco~\cite{cityhealth}. Meanwhile, greater Paris contains respectively the 2nd and 5th of the top five poorest and richest census tracks in all of France~\cite{iris}. Providing a solution to these challenges is of paramount importance to fulfill the economic and social promises that cities hold and to avoid them becoming sources of social and political instability.

The successful development and deployment of urban solutions to address these issues requires however both spatially fine-grained socioeconomic information as well as a detailed understanding on how wealth and the underlying urban topology are entangled. While this question has been thoroughly addressed before in previous urban sustainability studies ~\cite{seto2003modeling,stead2001relationships,mirmoghtadaee2012relationship,kinzig2005effects}, the socioeconomic maps used were generally coarse-grained (roughly in the order of $\text{km}^2$); the finest level of description being of course that provided by the census, thus hindering any detailed analysis. To cope with this, other works have relied upon socioeconomic proxies, mostly derived from large-scale digital datasets, to propose accurate and highly detailed socioeconomic status estimations. These include  communication patterns in call detail records~\cite{blumenstock} and social media~\cite{moro} or even restaurant data~\cite{ratti}. Most notably, some works have relied heavily upon large collections of satellite and street imagery to train deep learning models to predict wealth either from visual features~\cite{esra} or by predicting features known to correlate with wealth (such as night time light intensities~\cite{jean_neal} or models of car in census tracks~\cite{gebru2017}). Surprisingly, despite the overall use and reliability of these methods, no previous work has looked consistently at the features learned by these models, nor tried to uncover any existing correlations between the existing urban topology of the city and the high resolution socioeconomic map the model is tasked with predicting. Our aim is to close this gap by training a deep learning model to predict the socioeconomic status of a given location from its aerial image and in turn interpret its activation maps in terms of the underlying urban topology. 

More precisely, we first overlay three publicly available datasets, providing a complete description of five French cities in terms of socioeconomic and land use data, as well as aerial imagery. Subsequently, after merging the aerial imagery with the corresponding socioeconomic maps, we train a Convolutional Neural Network (CNN) model, to predict accurately the socioeconomic status of inhabited tiles. Next, by relying upon a gradient-weighted class activation mapping (Grad-CAM) for computing attribution maps ~\cite{gradcam,dementia_gradcam,alzheimer_gradcam}, we generate high resolution class discriminative activation maps, which are projected back onto the original image and overlaid with land use data. We thus generate empirical statistics on the features used by our model to predict socioeconomic status in terms of land use classes. This framework enables the inference of socioeconomic status at scales rarely seen before, while also indicating precisely the predictive features contained in the actual urban environment. Furthermore, it allows for the observation of distinct city-to-city patterns of correlations between urban topology and the distribution of wealth. In doing so, we find that when inferring socioeconomic status, our CNN models disregard existing correlations between land use and socioeconomic data and focus mostly on features contained within residential areas. The observation of these correlations is a milestone in ensuring the usability of these methods for policy makers, as they not only provide an enhanced understanding on the models' inner workings but they also may yield valuable insights into the urban development and planning of economically deprived areas within the city.

In what follows, we introduce our method by providing a concise overview of the used datasets and the architecture of the model we tasked with predicting socioeconomic status. We then describe how we used GradCAM to generate visual explanations from our model by means of activation maps. This in turn enabled us to explore  the predictions yielded by our network and observe how the inferred socioeconomic maps and the underlying urban patterns are correlated. We close with a discussion of the potential our method holds.

\section{Results}

\subsection*{Aerial, socioeconomic, land use datasets and their combination}

\begin{figure}[htbp]
\centerline{\includegraphics[width=.75\columnwidth]{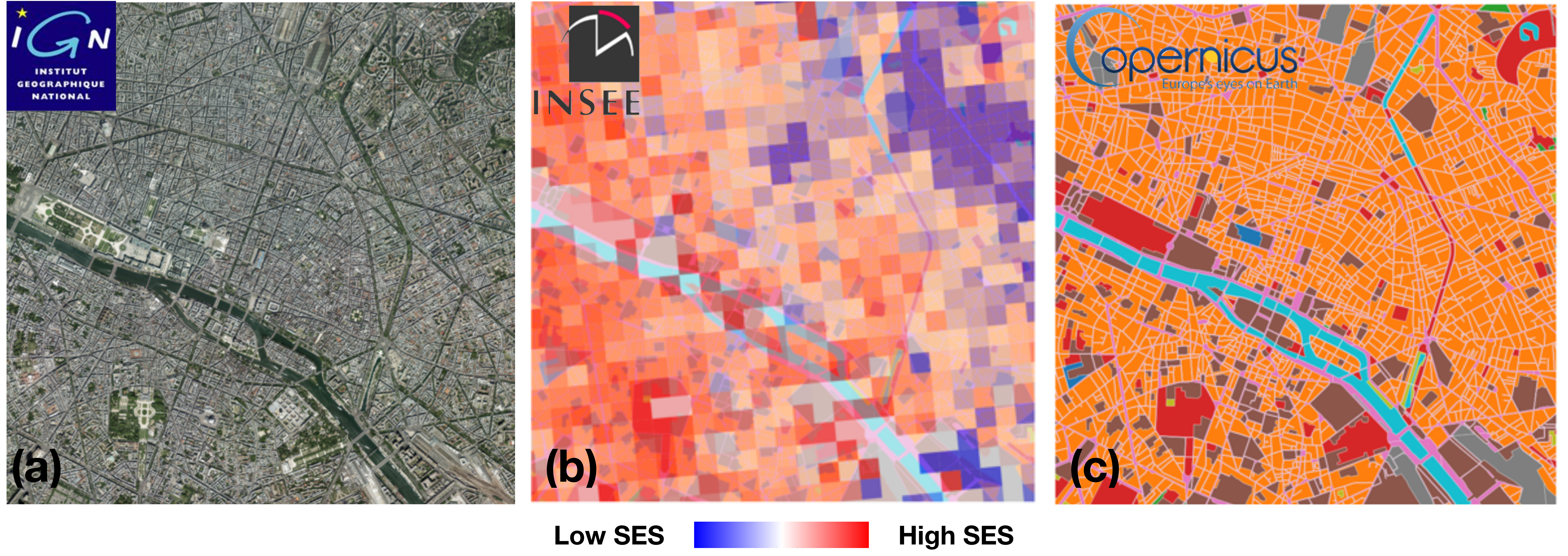}}
\caption{Sample of overlaid datasets (Paris): \textbf{(a)} 5km x 5km aerial tile (20cm/pixel); \textbf{(b)} Spatial distribution of income: each patch corresponds to a single 200m x 200m area with precise income data; \textbf{(c)} Land cover map of the same area, where each color represents a different urban class (not shown here for clarity).}
\label{fig:dataset}
\end{figure}

The fine-grained prediction of socioeconomic status from remotely sensed images requires both a detailed socioeconomic map of the area of interest and a dataset of aerial images provided at a sufficiently high resolution for the CNN to be able to detect localized features. Throughout this study, we rely on high-resolution \emph{aerial imagery} taken between 2013 and 2016 of the complete French metropolitan territory (at 20cm/pixel) issued by the National Geographical Information Institute (IGN)~\cite{ign}. This orthophotography dataset is distributed as a series of georeferenced 5km x 5km tiles provided at the department (administrative delimitation) level as shown in Figure~\ref{fig:dataset}a. This collection is then complemented with \emph{socioeconomic data} shared by the French National Institute of Statistics and Economic Studies (INSEE) in 2019~\cite{insee}. This corpus describes the winsorised average household income, estimated from the 2015 tax return in France, for each 4 hectare (200m x 200m) square patch (referred to henceforth as \emph{census cell}) across the whole French territory (Figure~\ref{fig:dataset}b). The combination of the socioeconomic map and the aerial data requires us to identify the aerial image (referred to henceforth as \emph{aerial tile}) corresponding to each 200m x 200m census cell, therefore enabling us to obtain for every socioeconomic patch its corresponding aerial view. For details regarding the winsorisation of the census data or its spatial matching with aerial images see the Supplementary Information (SI).

Next, to study the correlations between the distribution of wealth and the urban patterns, we gather \emph{land use data} shared by the European Environment Agency through the 2012 European Union Urban Atlas project. This dataset provides high-resolution land cover maps of roughly 700 urban areas of more than 100,000 inhabitants in EU28 and EFTA countries. It does so by segmenting each city into detailed polygons categorised into one of 27 standardized land use classes. In our study, only 19 classes were used, as infrequent and similar classes were respectively discarded and merged. Further information about the generation of this dataset can be found in the EU Urban Atlas~\cite{ua2012} and explained in the Materials section.

Following the spatial combination of these datasets, we single out five major cities located within the French metropolitan territory for further analysis. These were {Paris}, {Lyon}, {Marseille}, {Nice} and {Lille}. Geographical boundaries for each city/urban area are obtained from the ensuing land use data. For each city, income values from the socioeconomic patches are partitioned into one of five ($n_{SES}=5$) socioeconomic classes defined by the five quantiles of the city-wise income distribution, so that classes one and five correspond respectively to the bottom and top 20\% of earners in each city. In doing so, we obtain the spatial distribution of income specific to each city, as is shown in Figure~\ref{fig:dist_paris}a for Paris, where each colored pixel corresponds to a single socioeconomic patch. Similar results for the other cities can be found in the SI together with a detailed description of the used datasets. Note that all the collected datasets can be openly accessed and emanate from a single four-year time window hence minimizing their temporal misalignment.

\subsection*{Inferring socioeconomic status in cities}

To estimate the socioeconomic status of a given location we modify and train an  EfficientNetB0~\cite{efficientnet} model on socioeconomic and aerial tile pairs to predict the former from the latter. The trained model takes individual aerial tiles $s\in S$ set of tiles, as input and predicts a socioeconomic status (SES) label $\hat{y}(s)\in [1,...,n_{SES}]$ for each of them. This model was chosen over more classical models (like VGGs or ResNets) mostly for its architectural design: by making use of an effective compound coefficient to scale up CNNs, EfficientNets have been shown to achieve much better accuracy and efficiency than other convolutional neural networks. This enabled us to use the original fine-grained resolution of the aerial imagery as input to predict the SES at a lower computational cost. The architecture and parametrisation of our model are further described in the Methods section.

\begin{figure}[htbp]
\centerline{\includegraphics[width=.85\columnwidth]{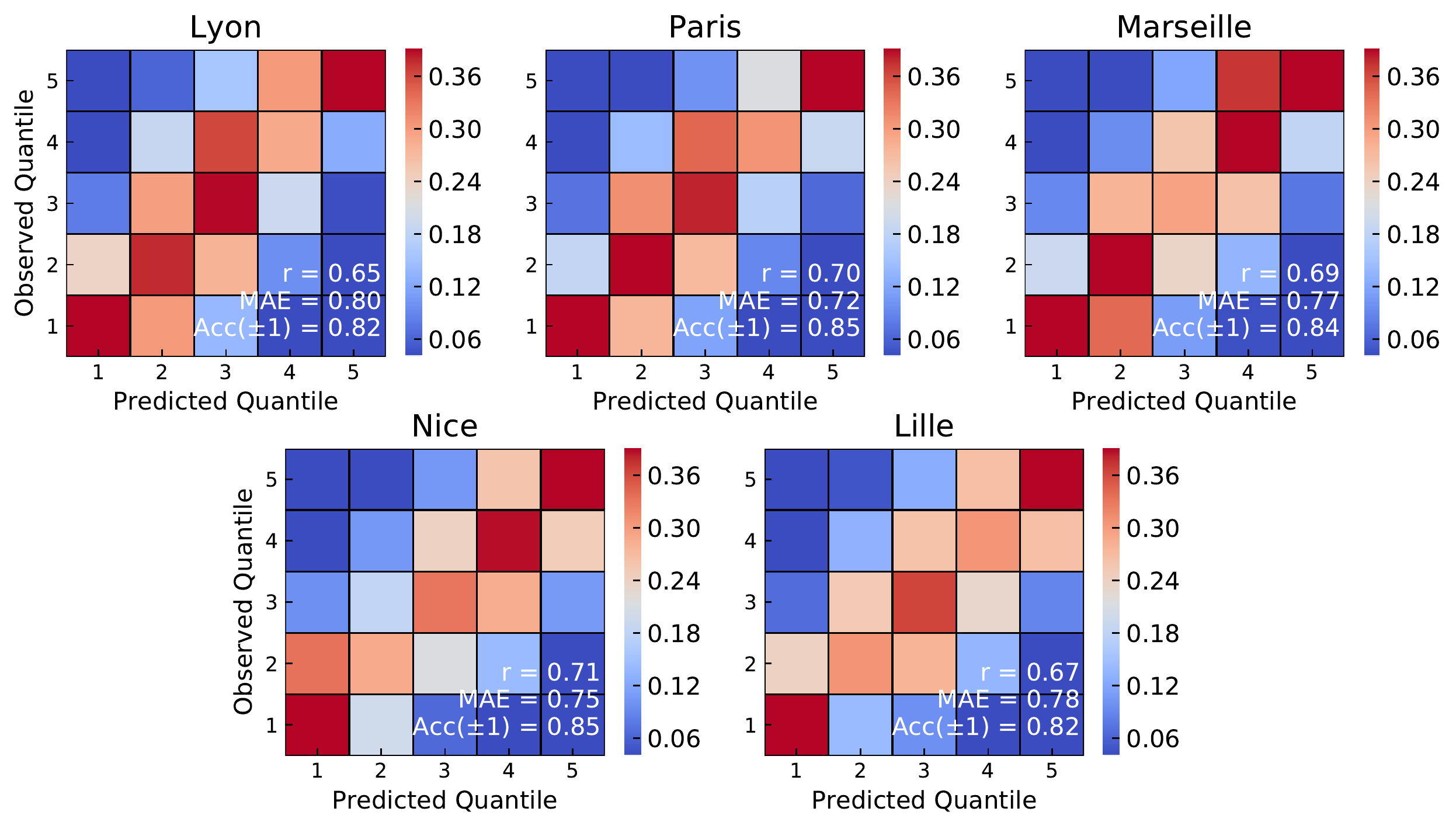}}
\caption{Observed performance of models trained to predict wealth in French cities: Confusion matrices between predicted and observed SES classes. In each plot, Pearson correlation coefficient ($r, p<0.01$) as well as mean average error (MAE) and accuracy within $\pm1$ classes are provided. All results are averaged over the $k$ cross-validation folds (with $k=5$). }
\label{fig_heatmaps}
\end{figure}

The modified EfficientNetB0 models are separately trained on each city to predict five-class socioeconomic status from 200m x 200m aerial tiles. To assess the quality of the predictions, we computed the metrics proposed by Suel et al.~\cite{esra} for each city in our dataset. We report the five-fold averaged cross-validation performance by means of separate confusion matrices in Figure \ref{fig_heatmaps}.

In each case, perfect prediction would correspond to a red diagonal and blue off-diagonal cells, while the greater the deviation from this pattern the greater the classification error made. We also report the $r$ Pearson correlation coefficient, mean average errors (MAE) and accuracy within one class (Acc) values between the observed and predicted SES classes in the figure keys. All values were computed for samples not seen either during training or validation of the model.

The quality of the predictions tends to be quite similar between cities with the average Pearson correlation between true and predicted quantiles varying between 0.65 for Lyon and 0.71 for Nice. Furthermore, in all cities, the mean average error between predicted and observed socioeconomic classes is bounded between 0.72 for Paris and 0.80 for Lyon. Although some misclassifications occur between neighboring quantiles, all tiles are classified within $\pm 1$ of their class with similar accuracies to those reported in previous studies (ranging from 0.82 to 0.85)~\cite{esra}. Interestingly, the classes predicted by our model with the lowest accuracy tends to vary between cities. For instance, in Paris and Lyon, the fourth quantile appears to be the hardest to predict correctly, while the second and third seem to be most prone to errors in Nice and Marseille respectively.
\begin{figure}[tbp!]
\centerline{\includegraphics[width=\columnwidth]{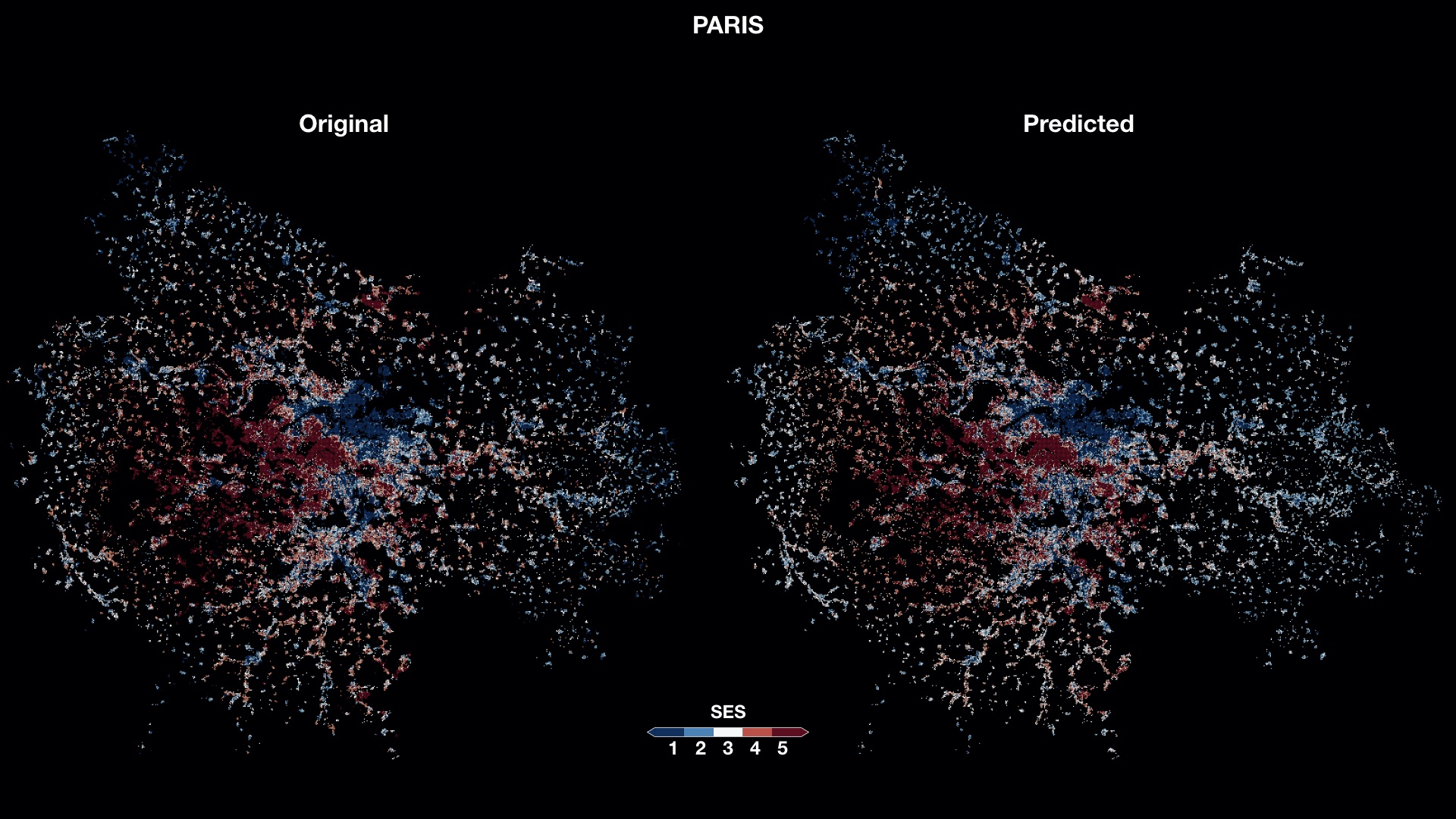}}
\caption{Maps of observed and predicted average income for Paris. Each pixel represents a single 200m x 200m tile with color indicating the average socioeconomic status of its inhabitants. Maps for the other cities can be found in the SI.}
\label{fig:dist_paris}
\end{figure}

To clearly visualise our predictions, we also depict the observed and predicted spatial distributions of income for Paris in Figure~\ref{fig:dist_paris} (and for all other cities in the SI). Contrary to other studies~\cite{esra,gebru2017},  each map pixel represents here a single cell corresponding to a unique tile, socioeconomic class pair. Consequently, predicted values are not aggregated but individually computed. It is therefore interesting to see that neighboring tiles of equal SES are overall correctly predicted to be of the same socioeconomic class, hence recovering the small-scale spatial homogeneity of the original income distribution (see center of Paris, Lyon or south-west of Nice and Marseille). At the same time, sub-urban regions characterized by socioeconomic disparities seem to be recovered. This can be seen in the predicted socioeconomic segregation between downtown Paris and its suburbs (in Figure~\ref{fig:dist_paris}), the west of Marseille, or the East of Lille (shown in SI) which further demonstrates that the trained models were able to recover the complex spatial distributions of income featured in our dataset with remarkable resolution and accuracy.

\subsection*{GradCAM derived model correlations between SES and urban topology}\label{formulation}

\begin{figure}[htbp]
\centerline{\includegraphics[width=0.7\columnwidth]{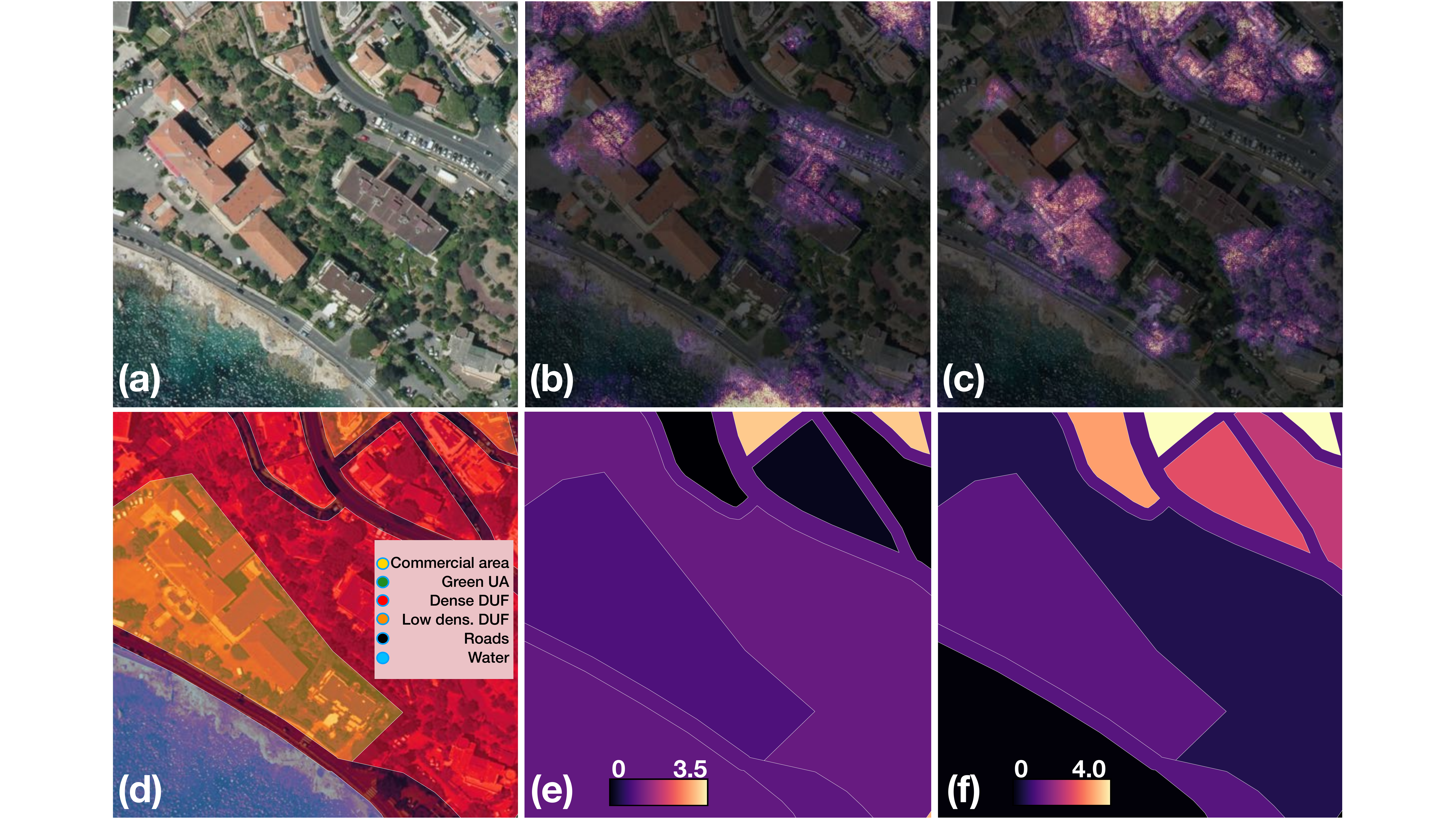}}
\caption{Model interpretability studies using Guided Grad-CAM (GGC). From an aerial tile (\textbf{a}), GGC is used to compute activation maps for the poorest (\textbf{b}) and wealthiest (\textbf{c}) socioeconomical class. The activation maps are then overlaid with the tile's tesselation into urban classes polygon (\textbf{d}) to compute the normalized ratio of activations per polygon for the poorest (\textbf{e}) and wealthiest (\textbf{f}) class. }
\label{fig:gradcam}
\end{figure}

Once all models are trained, we seek to identify the most salient features used by the CNNs to draw their predictions in terms of land use. This is a highly non-trivial task as it requires mapping the abstract activity patterns of the CNN back onto the original georeferenced space. To solve this, we apply guided gradient-weighted class activation mapping~\cite{gradcam} (Guided Grad-CAM) to identify visual features underlying the model’s predictions. Previous uses of this method include the identification of patterns learned by CNN models trained to predict Alzheimer's disease~\cite{dementia_gradcam,alzheimer_gradcam}, lung cancer~\cite{gradcam_cancer} as well as to classify wildlife~\cite{gradcam_animals} and plants diseases ~\cite{gradcam_plants}. Guided Grad-CAM follows the CNN’s gradient flow from individual output classes back onto the original image tile to establish an activation map, highlighting the input features most relevant to each class prediction. Figure~\ref{fig:gradcam} shows how Guided Grad-CAM (white features on dark background) is applied to a given aerial tile (panel (a)). The overlaid activation maps highlight the visual features that most trigger the prediction for the poorest (panel (b)) and richest (panel (c)) classes. These activation patterns can then be associated to the corresponding land use maps shown in Figure~\ref{fig:gradcam}d.

In turn, one can compute the overall activation values for different urban class polygons, and see which urban structures are the most salient when predicting the poorest (panel (e)) and richest classes (panel (f)). In what follows, we build upon this idea by first providing a clear definition on how to measure correlations between saliency maps and urban classes and then probing the model's learned features by evaluating the following two hypotheses:
\begin{itemize}
\item{\textbf{H0}: All urban classes contain features equally contributing to the model's prediction for the top and bottom SES class.}
\item{\textbf{H1}: Urban class activation maps are pairwise independent, i.e, activations for a given urban class are on average invariant to activation values for other urban classes.}
\end{itemize}
In what follows, we treat Guided-GradCAM maps as saliency maps describing pixels that are most determinant in the model's prediction. In doing so, every pixel $(i,j)$ in the saliency map of a given tile $s\in S$ is described both by its Guided GradCAM activation value $A_{ij}^s$ and the urban class $u\in U$ set of urban classes, of the polygon it belongs to via the mapping $f$ so that $\forall (i,j),\text{ } f(i,j)\in U$. Furthermore, we define the total activation for urban class $u\in U$ in tile $s\in S$ to be $ A_u^s = \sum_{(i,j)|f(i,j)=u}A_{i,j}$ so that each sample $s\in S$ is characterized by the tuple $(A_u^s)_{u\in U}$ where $A_u^s=\text{\textbackslash NA}$ for all urban classes $u\in U$ not contained in tile $s$. 

\subsubsection*{Univariate Correlations}
In order to examine the first hypothesis, we introduce the activation ratio for urban class $u$ defined as the ratio between the sum of activations and the expected sum of activations in a random diffusion model concentrated by polygons of urban class $u$:
$$r_u^s =  \frac{A_u^s}{F_u\sum_{k\in U}A_k^s}$$
$\text{ with }F_u\text{ the fraction of area occupied by polygons of class }$. This metric actually describes how the pixels that are triggered the most for the prediction of a socioeconomic class are distributed across urban classes. For instance, if no urban class contains more important features than others (\textbf{H0}) its expected value should be equal to one over all samples and for all urban classes contained in each sample. To obtain an aggregated view on this metric, we also computed the expected activation ratio for urban class $u$, $x_u^{\sigma}=\mathbb{E}_{\sigma}(r_u^s)$ on the set $\sigma$ of aerial tiles predicted to be of \emph{low} ($\sigma_{\text{low}}=\{s\in S|\hat{y}_s\in \{1,2\}\}$) or \emph{high} ($\sigma_{\text{high}}=\{s\in S|\hat{y}_s\in \{3,4\}\}$) SES. Further, to provide a more grounded comparison to the actual link between SES and urban patterns, we measured the empirical probabilities $\hat{p}_u^{\text{low}}$ and $\hat{p}_u^{\text{high}}$ that an aerial tile is 
of low or high SES given that it contains urban class $u$.

\begin{figure}[h]
\centerline{\includegraphics[width=.85\columnwidth]{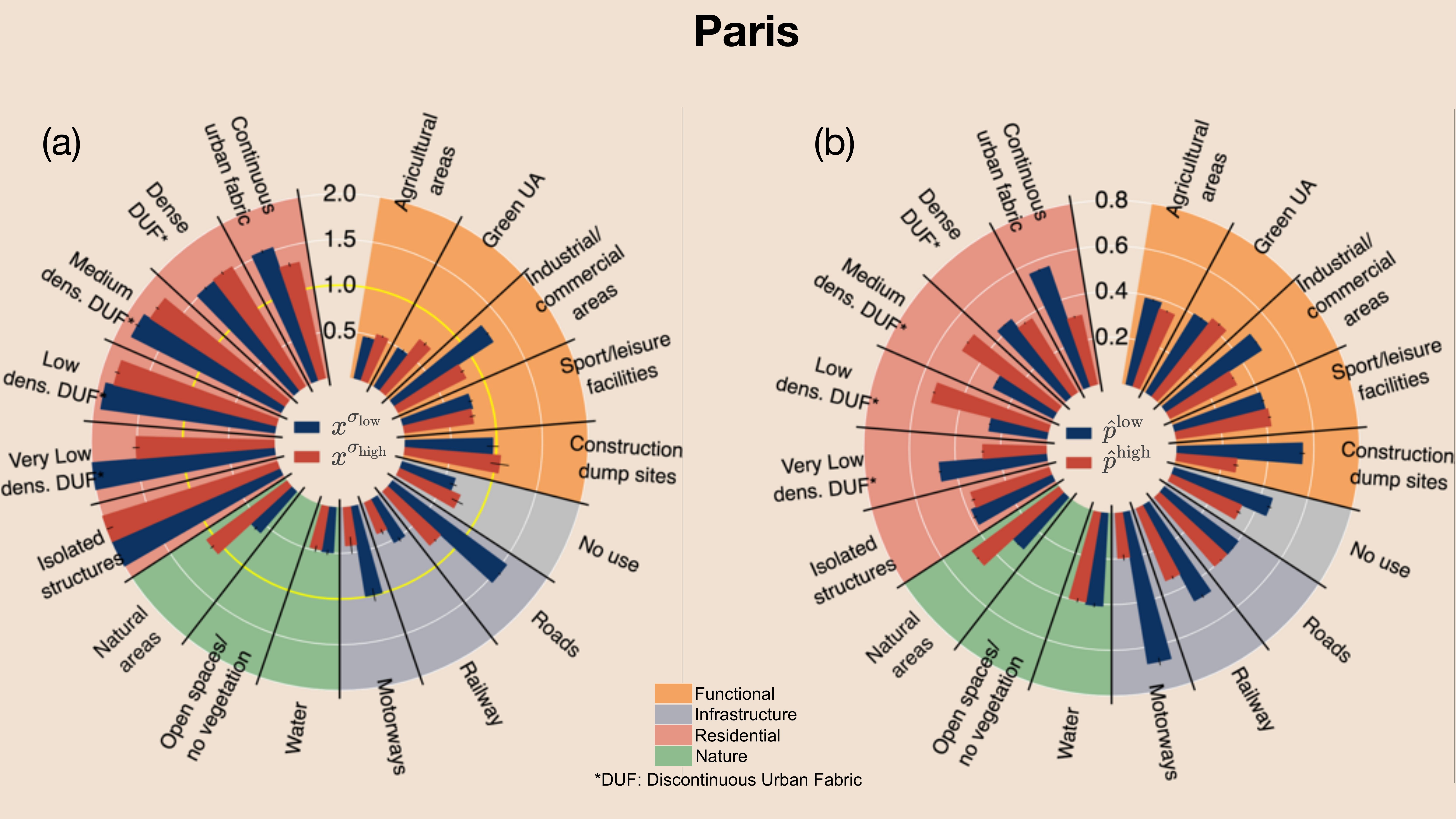}}
\caption{Correlations between urban topology and socioeconomic status in the city of Paris: (a) Mean model activation rate per urban class with bootstrapped 95\% confidence interval for samples predicted as respectively as low SES (blue) or high SES (red) by the model. (b) Estimated probability of an urban polygon belonging to the bottom or top quintile of the income distribution with bootstrapped  95\% confidence interval.}
\label{fig:gradcam_paris}
\end{figure}

We now identify the salient visual features underlying the model’s predictions in terms of land use for samples predicted to be of low and high SES. In this analysis, we refer to an urban class $u$ being over-activated if its expected activation ratio is greater than one, i.e $x_u^{\sigma} > 1$. Conversely, it is referred to as under-activated whenever its expected activation ratio is below one, i.e $x_u^\sigma < 1$. Results for Paris are shown in Figure~\ref{fig:gradcam_paris} while other cities are included in the SI. To ease the description in Figure~\ref{fig:gradcam_paris}, we group each urban class into four categories as functional, infrastructure, nature, or residential areas.

Looking at the spectrum of activations, certain structural elements tend to be shared across all trained models. Residential areas are over-activated both when predicting low and high SES, with  some differences remaining between cities: for models trained to predict SES in Lyon, Lille and Marseille, residential areas tend to activate to a greater extent when predicting high SES, while in Paris and Nice, residential areas appear over-activated when predicting low SES.

Infrastructural areas on the other hand follow quite a different pattern as they tend to be over-activated for low SES class predictions. For example, railways and motorways are triggered in Lyon and Marseille, while features contained within roads are generally active for all cities.

More generally, the analysis of the univariate correlations is evidence that CNN models trained to predict SES in urban areas are mostly reliant on features lying within residential areas to draw their predictions, while the degree to which other amenities are determinant in the prediction shifts from city to city, therefore invalidating \textbf{(H0)}.

Significantly different patterns are observed when looking at the empirical probabilities, $\hat{p}_u^{\text{low}}$ and $\hat{p}_u^{\text{high}}$ for Paris as shown in Figure~\ref{fig:gradcam_paris}b (and for other cities in the SI). For instance, while the type of residential areas contained in a given tile greatly varies between SES classes, low and medium density areas are more prone to being located within high SES tiles in any city while very high density areas are more likely to be contained within low SES tiles. Furthermore, interesting patterns can be observed for other categories. For instance, tiles containing motorways, railways and commercial/industrial units were mostly associated to low SES areas whereas those including natural areas were more likely to be of high SES.

\subsubsection*{Bivariate Correlations}
We have so far established that the models we train to estimate SES derive their predictions mostly from features contained within residential areas. Even though the previous analysis sheds light on the features behind the models prediction for both low and high SES, it is still unclear how the importance of a residential area is affected by the fact it neighbors other types of urban classes. In order to address this question, we mimic the former analysis and measure the expected co-activation ratio for pairs of urban classes $u$ and $v$, as $c_{u,v}^{\sigma}=\mathbb{E}_{\sigma'}(r_u^s)/\mathbb{E}_{\sigma}(r_u^s)$. Here $\sigma'=\{ s\in \sigma | A_u^s,A_v^s\neq \text{\textbackslash NA}\}$ is the set of aerial tiles where urban class $v$ is present alongside urban class $u$, thus $\sigma' \subseteq \sigma$. Therefore for a given urban class $u$ and for any other urban class $v$, if \textbf{(H1)} were to hold, the expected value of the activation ratio $\mathbb{E}_{\sigma'}(r_u^s)$ should remain relatively unchanged thus yielding a value for $c_{u,v}^{\sigma}$ close to 1.

\begin{figure}[htbp]
\centerline{\includegraphics[width=.85\columnwidth]{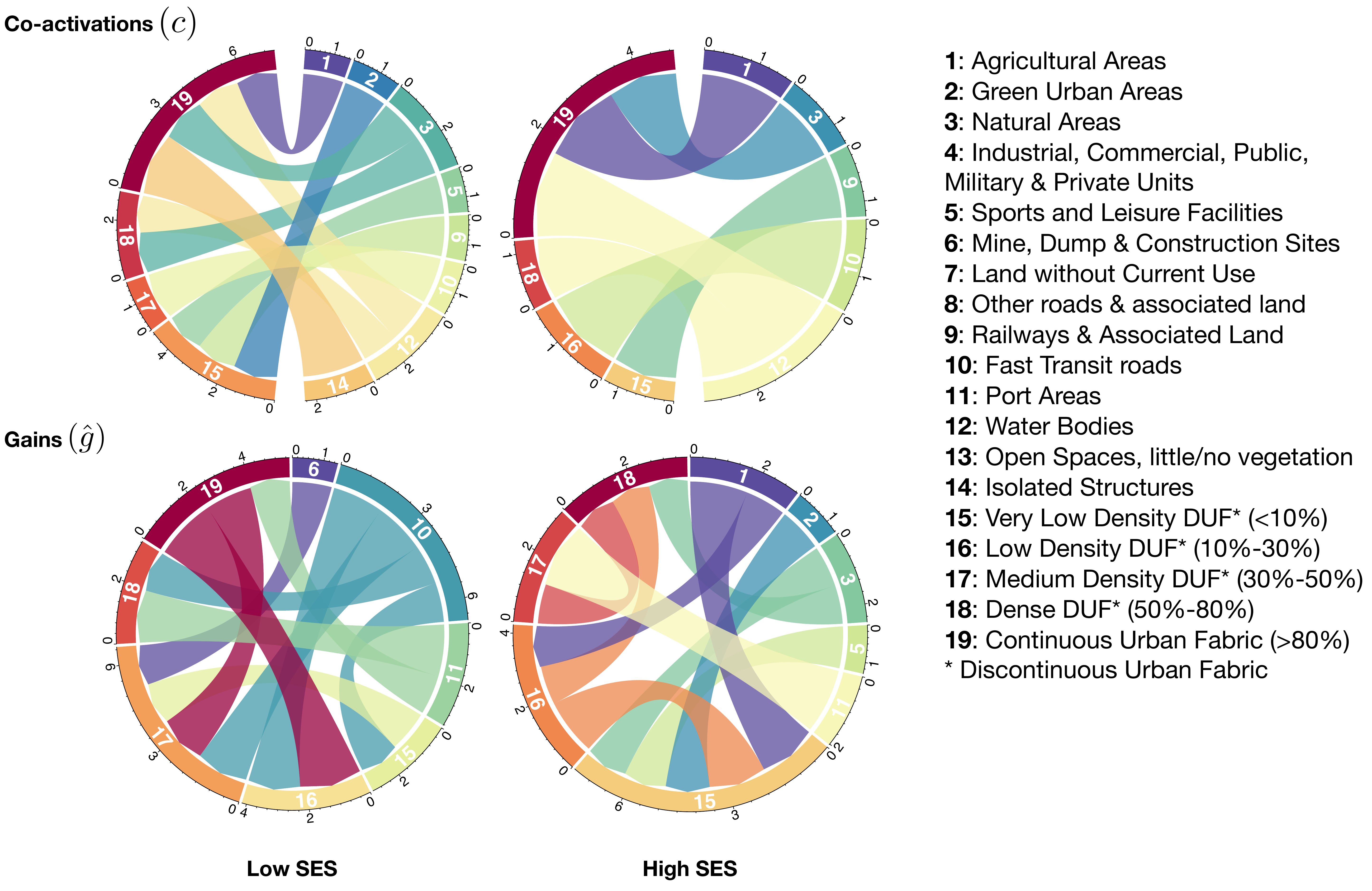}}
\caption{Chord diagrams of coactivations (top) and coappearance gains (bottom) estimated for both low (left) and high (right) SES in the city of Paris. Whenever urban class $i$ appears more activated or is more likely to be of a given SES in the presence of urban class $j$ than when appearing by itself or any other random urban class, it is represented in the diagram as $j\rightarrow i$. Each circular segment represents an urban class $i$ connected by chords with width equal to the coactivation/coappearance gain. Colors are assigned in order of appearance for better visualization based on a common colormap and are not exclusive to a given class. For simplicity, we illustrate only pairs of classes with co-activations and coappearance gains values greater than 1.2 (at least 20\% improvement over the univariate case). Among these only the top ten ranked by either coactivations and coappearance gains are shown.}
\label{fig:paris_res}
\end{figure}

Additionally, in order to provide a grounded comparison to the actual data, we compute the co-appearance gains $\hat{g}^{\text{low}}_{(u,v)}=\hat{p}^{\text{low}}_{u,v}/\hat{p}^{\text{low}}_{u}$ and $\hat{g}^{\text{high}}_{(u,v)}=\hat{p}^{\text{high}}_{u,v}/\hat{p}^{\text{high}}_{u}$ respectively describing how much more/less likely an urban polygon of class $u\in U$ is to be within a \emph{low} or \emph{high} SES tile given that it co-occurs with an urban polygon of class $v\in U$. Notice that for a pair of urban classes $(u,v)$ neither co-activation ratios nor co-appearance gains are symmetrical. As we focus on residential areas, we constrain $u$ to be from a very low to very high density residential area (while omitting isolated residential areas since they are, by definition, less likely to co-occur with other urban classes). Co-activations and co-appearance gains for the city of Paris are shown in Figure~\ref{fig:paris_res} while results for other cities may be found in the SI. As before, we refer to an urban class $u$ being over-coactivated in the presence of class $v$ whenever its coactivation ratio is greater than one, i.e $c_{u,v}^{\sigma}>1$.  Additionally, in order to ease the analysis of the results, we only report pairs of urban classes ($u,v$) where an increase of over 20\% of the baseline expected activation ratio or probability is observed.  

Regarding the co-activations (see definition above), most residential areas appear over-coactivated when they neighbor clearly non-residential areas. For instance, with the exception of low SES predictions in Marseille and Nice, very high and high density residential areas are over-coactivated when occurring with natural areas or water bodies. This also extends to agricultural areas in the case of Paris and Lille. Interestingly, these results are observed both for high and low SES predictions. Very low density areas on the other hand appear over-coactivated for low SES predictions whenever they neighbor green urban areas or leisure facilities in Paris, Lyon and  Marseille. Focusing solely on residential areas, with the exception of low SES predictions in Marseille and Nice, we observe that whenever residential areas of different densities co-occur with one another, no over-coactivation is observed. This can be seen in the absence of chords linking urban residential areas in the aforementioned cases in Figure~\ref{fig:paris_res} (top) and in the SI. We can therefore conclude, that urban classes activate differently depending on the other classes they co-occur with, thus disproving \textbf{(H1)}. For residential areas, these differences appear when they co-occur with non-residential ones, in which case they seem to be over-coactivated implying that our models tend to disregard pixels contained in these non-residential urban classes in favour of those within residential areas.

This result is quite surprising especially when we consider that the amenities that surround a given residential area are actually quite determinant in terms of the socioeconomic status of the people inhabiting it. For instance, when looking at the co-appearance gains in Figure~\ref{fig:paris_res} (bottom) and in the SI, we see that low and medium density residential areas co-occurring next to railways and fast transit roads are more likely to be of low SES than when co-occurring with other urban classes. Similarly, very high density residential areas neighboring water bodies or natural areas are more likely to be of high SES than otherwise.

\section*{Discussion}

The successful deployment of urban policies aiming to curbing social problems like income inequality, segregation and poverty requires an updated and up-scaled socioeconomic description of the city. While nation-wide censuses are meant to provide such information, their prohibitive cost makes their collection rather infrequent. This in turn induces a lag between the actual socioeconomic changes and when they are recorded in the data. As new sources of remotely sensed imagery become available, the opportunity to frequently complement the census with socioeconomic information inferred from these digital collections has never been greater. This has been recognised even by policy makers who rely on these novel solutions more and more despite their inherit lack of interpretability.

Our aim in this study has been to address this shortfall by 1) building a dataset from open data sources ready to be used for deep learning based urban solutions 2) providing a CNN-based framework for predicting socioeconomic status from aerial imagery for five French cities, 3) establishing a method tointerpret the learned activation patterns to infer SES by mapping them back onto the original aereal images, and 4) examining the activations derived from these models in terms of land cover. In doing so, we have built a model able to generate socioeconomic predictions with state-of-the-art performance and high spatial resolution. More importantly, we have here provided a way to interpret the activation patterns of trained CNN models in terms of land use. Results showed that models trained to infer SES from aerial imagery seem to rely on features contained mostly within residential areas rather than non-residential ones. Finally, we have also observed that this pattern of activations differs significantly from the statistical information one can derive by joining land cover and socioeconomic information.

Nevertheless, the interpretation of activations provided here has to be taken with caution as it still remains unclear how the choice of architecture, the training parameters or the transferability of the learned patterns in space and time change the observed correlations. These caveats indicate though the research avenues further work needs to follow as different models, resolutions and interpretability techniques need to be explored to reach the end-goal of deploying these models with a more complete understanding of their inner mechanisms. To help this end, we make all our code and data pointers openly available to the community (see Data Availability and Code Availability Statements). We hope this proof of concept motivates further work in this field, where the identification of global patterns of poverty could have far-reaching impact by enabling the swift allocation of needed resources to the most vulnerable.

\section*{Methods}

\subsection*{Model description for SES inference}

\begin{figure}[htbp]
\centerline{\includegraphics[width=\columnwidth]{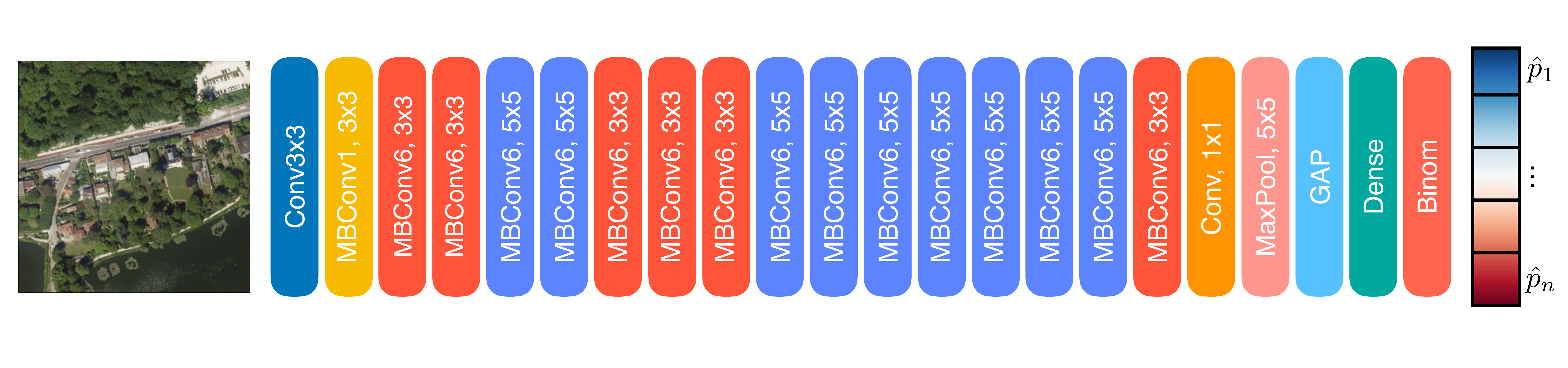}}
\caption{Model Architecture: The model takes the aerial tile as an input which is then fed through several MBConv blocks. The feature maps end up going through a global average pooling layer and a dense layer to output a single value $p$. From it, probabilities for each socioeconomic class are generated from a binomial distribution $B{(N_{SES},p)}$}
\label{fig:archi}
\end{figure}

Our model used the original EfficientNetB0 (EB0) architecture consisting of a series of mobile inverted bottleneck blocks (MBConv) augmented with squeeze-and-excitation optimisation. In addition we added a max pooling layer, to downsample the feature maps, followed by global average pooling and dense layers (see Figure~\ref{fig:archi}). In order to yield an estimated probability of income quantiles for each input, we followed the approach used by Suel \textit{et al.}~\cite{esra}, and connected the dense layer to a binomial layer taking a single $p$ value, interpreted as a probability with which Bernoulli trials are performed, as input and outputting the probabilities for each SES class $Y=[\hat{y}_1, ..., \hat{y}_{n_{SES}}]$, with:
$$\hat{y}_k = \binom{n_{SES}}{k-1}p^{k-1}(1-p)^{n_{SES}-(k-1)},\quad k\in[1, ..., n_{SES}]$$
In doing so, the model becomes sensitive to the ordinal relationship existing between quantiles. Training is then conducted with a classic categorical cross-entropy loss. 

It is worth mentioning that the original EB0 model was trained on object-centric 224 x 224 x 3 images and therefore its initial weights were  unfit for the dataset we wanted to apply it onto. Even though the CNN's weights were initially set to those of a model trained on ImageNet (except for the final added layers which are initialised randomly), we didn't freeze any of the model's layers to train the model from scratch. This of course assumes that the optimisation procedure will more readily converge by adapting the weights from the optimal ones learned at lower resolutions than by actually learning them from scratch. This assumption is reasonable since the first layers of the network will already be sensitive to low-level features.

In all our experiments, we augmented the data through random horizontal and vertical flipping of the input images. Images were scaled to 800 x 800 x 3 to avoid decreasing excessively the original image resolution while keeping economic the use of our computational resources. An Adam variant of Stochastic Gradient Descent was used to learn the models weights. Initially, we started with a learning rate of $8.10^{-5}$ and decreased it by 90\% every 3 epochs until the validation loss showed no improvement for more than 10 epochs. All networks were trained on either NVIDIA Tesla K20X or V100 graphics processing units for at most 30 epochs with 2500 samples in each epoch. Furthermore, within each city, we used five-fold cross validation. In each fold, 80\% of data was used for training the network and the remaining 20\% were withheld. The training set was further randomly subdivided with a 75-25\% split for inner-fold training and validation. Every reported performance metric is then averaged over all folds.

\subsection*{Guided gradient-weight class activation maps}\label{formulation}

Guided Grad-CAM follows the CNN’s gradient flow from individual output classes back onto the original image tile to establish an activation map, highlighting the input features most relevant to each CNN prediction.  To generate such mappings, we first computed saliency maps via guided backpropagation, producing a pixel-space gradient map of predicted class scores with respect to pixel intensities of the input image. This gradient visualization, though fine-grained is known not to be class-discriminative. It can however be combined with Grad-CAM via pointwise multiplication to generate Guided Grad-CAM (see SI for more information). Grad-CAM itself results from the dot product of the feature map of the last convolutional layer and the partial derivatives of predicted class scores with respect to the neurons in the last convolutional layer. All mappings were done in a cluster of 96 CPUs.

\section*{Data Availability Statement}

In this paper we built on the combination of three publicly available datasets. One was issued by the National Geographical Information Institute (IGN) and contained aerial images about French municipals~\cite{ign}. The second was provided by the French National Institute of Statistics and Economic Studies (INSEE) in 2019~\cite{insee} and provided a high resolution socioeconomic map. The third was shared by the European Environment Agency through the 2012 European Union Urban Atlas project about the EU28 and EFTA countries~\cite{ua2012}. All datasets are public and openly accessible online. Figures depicting raw data are Figure~\ref{fig:dataset}, \ref{fig:dist_paris}, \ref{fig:gradcam}, and \ref{fig:archi}.

\section*{Code Availability Statement}
Code developed for the actual research is shared through an open repository~\cite{github_jaklevab} along some sample data. For any further question please contact the corresponding authors.


\begin{thebibliography}{10}
\urlstyle{rm}
\expandafter\ifx\csname url\endcsname\relax
  \def\url#1{\texttt{#1}}\fi
\expandafter\ifx\csname urlprefix\endcsname\relax\def\urlprefix{URL }\fi
\expandafter\ifx\csname doiprefix\endcsname\relax\def\doiprefix{DOI: }\fi
\providecommand{\bibinfo}[2]{#2}
\providecommand{\eprint}[2][]{\url{#2}}

\bibitem{united20182018}
\bibinfo{author}{Nations, U.}
\newblock \bibinfo{title}{2018 revision of world urbanization prospects}
  (\bibinfo{year}{2018}).

\bibitem{glaeser}
\bibinfo{author}{Glaeser, E.~L.} \& \bibinfo{author}{Joshi-Ghani, A.}
\newblock \emph{\bibinfo{title}{The urban imperative: towards competitive
  cities}} (\bibinfo{publisher}{Oxford University Press},
  \bibinfo{year}{2015}).

\bibitem{cityhealth}
\bibinfo{author}{Gourevitch, M.~N.}, \bibinfo{author}{Athens, J.~K.},
  \bibinfo{author}{Levine, S.~E.}, \bibinfo{author}{Kleiman, N.} \&
  \bibinfo{author}{Thorpe, L.~E.}
\newblock \bibinfo{journal}{\bibinfo{title}{City-level measures of health,
  health determinants, and equity to foster population health improvement: The
  city health dashboard}}.
\newblock {\emph{\JournalTitle{American Journal of Public Health}}}
  \textbf{\bibinfo{volume}{109}}, \bibinfo{pages}{585--592},
  \doiprefix\url{10.2105/AJPH.2018.304903} (\bibinfo{year}{2019}).
\newblock \bibinfo{note}{PMID: 30789770},
  \eprint{https://doi.org/10.2105/AJPH.2018.304903}.

\bibitem{iris}
\bibinfo{author}{INSEE}.
\newblock \bibinfo{title}{Revenus, pauvret\'e et niveau de vie en 2014.}
  (\bibinfo{year}{2014}).

\bibitem{seto2003modeling}
\bibinfo{author}{Seto, K.~C.} \& \bibinfo{author}{Kaufmann, R.~K.}
\newblock \bibinfo{journal}{\bibinfo{title}{Modeling the drivers of urban land
  use change in the pearl river delta, china: Integrating remote sensing with
  socioeconomic data}}.
\newblock {\emph{\JournalTitle{Land Economics}}} \textbf{\bibinfo{volume}{79}},
  \bibinfo{pages}{106--121} (\bibinfo{year}{2003}).

\bibitem{stead2001relationships}
\bibinfo{author}{Stead, D.}
\newblock \bibinfo{journal}{\bibinfo{title}{Relationships between land use,
  socioeconomic factors, and travel patterns in britain}}.
\newblock {\emph{\JournalTitle{Environment and Planning B: Planning and
  Design}}} \textbf{\bibinfo{volume}{28}}, \bibinfo{pages}{499--528}
  (\bibinfo{year}{2001}).

\bibitem{mirmoghtadaee2012relationship}
\bibinfo{author}{Mirmoghtadaee, M.}
\newblock \bibinfo{journal}{\bibinfo{title}{The relationship between land use,
  socio-economic characteristics of inhabitants and travel demand in new
  towns--a case study of hashtgerd new town (iran)}}.
\newblock {\emph{\JournalTitle{International Journal of Urban Sustainable
  Development}}} \textbf{\bibinfo{volume}{4}}, \bibinfo{pages}{39--62}
  (\bibinfo{year}{2012}).

\bibitem{kinzig2005effects}
\bibinfo{author}{Kinzig, A.~P.}, \bibinfo{author}{Warren, P.},
  \bibinfo{author}{Martin, C.}, \bibinfo{author}{Hope, D.} \&
  \bibinfo{author}{Katti, M.}
\newblock \bibinfo{journal}{\bibinfo{title}{The effects of human socioeconomic
  status and cultural characteristics on urban patterns of biodiversity}}.
\newblock {\emph{\JournalTitle{Ecology and Society}}}
  \textbf{\bibinfo{volume}{10}} (\bibinfo{year}{2005}).

\bibitem{blumenstock}
\bibinfo{author}{Blumenstock, J.~E.}
\newblock \bibinfo{journal}{\bibinfo{title}{Fighting poverty with data}}.
\newblock {\emph{\JournalTitle{Science}}} \textbf{\bibinfo{volume}{353}},
  \bibinfo{pages}{753--754}, \doiprefix\url{10.1126/science.aah5217}
  (\bibinfo{year}{2016}).
\newblock
  \eprint{https://science.sciencemag.org/content/353/6301/753.full.pdf}.

\bibitem{moro}
\bibinfo{author}{Llorente, A.}, \bibinfo{author}{Garcia-Herranz, M.},
  \bibinfo{author}{Cebrian, M.} \& \bibinfo{author}{Moro, E.}
\newblock \bibinfo{journal}{\bibinfo{title}{Social media fingerprints of
  unemployment}}.
\newblock {\emph{\JournalTitle{PLOS ONE}}} \textbf{\bibinfo{volume}{10}},
  \bibinfo{pages}{1--13}, \doiprefix\url{10.1371/journal.pone.0128692}
  (\bibinfo{year}{2015}).

\bibitem{ratti}
\bibinfo{author}{Dong, L.}, \bibinfo{author}{Ratti, C.} \&
  \bibinfo{author}{Zheng, S.}
\newblock \bibinfo{journal}{\bibinfo{title}{Predicting
  neighborhoods{\textquoteright} socioeconomic attributes using restaurant
  data}}.
\newblock {\emph{\JournalTitle{Proceedings of the National Academy of
  Sciences}}} \doiprefix\url{10.1073/pnas.1903064116} (\bibinfo{year}{2019}).
\newblock
  \eprint{https://www.pnas.org/content/early/2019/07/09/1903064116.full.pdf}.

\bibitem{esra}
\bibinfo{author}{Suel, E.}, \bibinfo{author}{Polak, J.~W.},
  \bibinfo{author}{Bennett, J.~E.} \& \bibinfo{author}{Ezzati, M.}
\newblock \bibinfo{journal}{\bibinfo{title}{Measuring social, environmental and
  health inequalities using deep learning and street imagery}}.
\newblock {\emph{\JournalTitle{Scientific Reports}}}
  \textbf{\bibinfo{volume}{9}}, \bibinfo{pages}{6229},
  \doiprefix\url{10.1038/s41598-019-42036-w} (\bibinfo{year}{2019}).

\bibitem{jean_neal}
\bibinfo{author}{Jean, c.} \emph{et~al.}
\newblock \bibinfo{journal}{\bibinfo{title}{Combining satellite imagery and
  machine learning to predict poverty}}.
\newblock {\emph{\JournalTitle{Science}}} \textbf{\bibinfo{volume}{353}},
  \bibinfo{pages}{790--794}, \doiprefix\url{10.1126/science.aaf7894}
  (\bibinfo{year}{2016}).
\newblock \eprint{http://science.sciencemag.org/content/353/6301/790.full.pdf}.

\bibitem{gebru2017}
\bibinfo{author}{\color{black}Timnit Gebru} \emph{et~al.}
\newblock \bibinfo{journal}{\bibinfo{title}{Using deep learning and google
  street view to estimate the demographic makeup of neighborhoods across the
  united states}}.
\newblock {\emph{\JournalTitle{Proceedings of the National Academy of
  Sciences}}} \textbf{\bibinfo{volume}{114}}, \bibinfo{pages}{13108--13113},
  \doiprefix\url{10.1073/pnas.1700035114} (\bibinfo{year}{2017}).

\bibitem{gradcam}
\bibinfo{author}{{Selvaraju}, R.~R.} \emph{et~al.}
\newblock \bibinfo{title}{Grad-cam: Visual explanations from deep networks via
  gradient-based localization}.
\newblock In \emph{\bibinfo{booktitle}{2017 IEEE International Conference on
  Computer Vision (ICCV)}}, \bibinfo{pages}{618--626},
  \doiprefix\url{10.1109/ICCV.2017.74} (\bibinfo{year}{2017}).

\bibitem{dementia_gradcam}
\bibinfo{author}{Iizuka, T.}, \bibinfo{author}{Fukasawa, M.} \&
  \bibinfo{author}{Kameyama, M.}
\newblock \bibinfo{journal}{\bibinfo{title}{Deep-learning-based
  imaging-classification identified cingulate island sign in dementia with lewy
  bodies}}.
\newblock {\emph{\JournalTitle{Scientific Reports}}}
  \textbf{\bibinfo{volume}{9}}, \bibinfo{pages}{8944},
  \doiprefix\url{10.1038/s41598-019-45415-5} (\bibinfo{year}{2019}).

\bibitem{alzheimer_gradcam}
\bibinfo{author}{Tang, Z.} \emph{et~al.}
\newblock \bibinfo{journal}{\bibinfo{title}{Interpretable classification of
  alzheimer's disease pathologies with a convolutional neural network
  pipeline}}.
\newblock {\emph{\JournalTitle{Nature Communications}}}
  \textbf{\bibinfo{volume}{10}}, \bibinfo{pages}{2173},
  \doiprefix\url{10.1038/s41467-019-10212-1} (\bibinfo{year}{2019}).

\bibitem{ign}
\bibinfo{author}{IGN}.
\newblock \bibinfo{title}{Donn\'ees \uppercase{ORTHO HR}}
  (\bibinfo{year}{2020}).
\newblock \bibinfo{note}{Data retrieved from,
  \url{https://geoservices.ign.fr/documentation/diffusion/telechargement-donnees-libres.html\#ortho-hr-sous-licence-ouverte}.
  Accessed the 01/03/2020.}

\bibitem{insee}
\bibinfo{author}{INSEE}.
\newblock \bibinfo{title}{Donn\'ees carroy\'ees} (\bibinfo{year}{2019}).
\newblock \bibinfo{note}{Data retrieved from the Filosofi 2015 gridded data,
  \url{https://www.insee.fr/fr/statistiques/4176290?sommaire=4176305}. Accessed
  the 01/03/2020.}

\bibitem{ua2012}
\bibinfo{author}{Service, C. L.~M.}
\newblock \bibinfo{title}{Urban atlas} (\bibinfo{year}{2012}).
\newblock \bibinfo{note}{Data retrieved from the France container,
  \url{https://land.copernicus.eu/local/urban-atlas/urban-atlas-2012}. Accessed
  the 01/03/2020.}

\bibitem{efficientnet}
\bibinfo{author}{Tan, M.} \& \bibinfo{author}{Le, Q.}
\newblock \bibinfo{title}{Efficientnet: Rethinking model scaling for
  convolutional neural networks}.
\newblock In \emph{\bibinfo{booktitle}{International Conference on Machine
  Learning}}, \bibinfo{pages}{6105--6114} (\bibinfo{year}{2019}).

\bibitem{gradcam_cancer}
\bibinfo{author}{Hosny, A.} \emph{et~al.}
\newblock \bibinfo{journal}{\bibinfo{title}{Deep learning for lung cancer
  prognostication: A retrospective multi-cohort radiomics study}}.
\newblock {\emph{\JournalTitle{PLOS Medicine}}} \textbf{\bibinfo{volume}{15}},
  \bibinfo{pages}{1--25}, \doiprefix\url{10.1371/journal.pmed.1002711}
  (\bibinfo{year}{2018}).

\bibitem{gradcam_animals}
\bibinfo{author}{Miao, Z.} \emph{et~al.}
\newblock \bibinfo{journal}{\bibinfo{title}{Insights and approaches using deep
  learning to classify wildlife}}.
\newblock {\emph{\JournalTitle{Scientific Reports}}}
  \textbf{\bibinfo{volume}{9}}, \bibinfo{pages}{8137},
  \doiprefix\url{10.1038/s41598-019-44565-w} (\bibinfo{year}{2019}).

\bibitem{gradcam_plants}
\bibinfo{author}{Toda, Y.} \& \bibinfo{author}{Okura, F.}
\newblock \bibinfo{journal}{\bibinfo{title}{How convolutional neural networks
  diagnose plant disease}}.
\newblock {\emph{\JournalTitle{Plant Phenomics}}}
  \textbf{\bibinfo{volume}{2019}}, \doiprefix\url{10.1155/2019/9237136}
  (\bibinfo{year}{2019}).

\bibitem{github_jaklevab}
\bibinfo{author}{Abitbol, J.~L.}
\newblock \bibinfo{title}{Coding repository}.
\newblock \bibinfo{note}{The implementation of our model and data preprocessing
  pipelines are available online
  at~\url{https://github.com/jaklevab/SESEfficientCAM/}}.

\bibitem{maps_price}
\bibinfo{author}{Techcrunch}.
\newblock \bibinfo{title}{Google revamps its google maps developer platform}
  (\bibinfo{year}{2018}).
\newblock
  \bibinfo{note}{\url{https://techcrunch.com/2018/05/02/google-revamps-its-google-maps-developer-platform/}}.

\bibitem{safety}
\bibinfo{author}{De~Nadai, M.} \emph{et~al.}
\newblock \bibinfo{title}{Are safer looking neighborhoods more lively?: A
  multimodal investigation into urban life}.
\newblock In \emph{\bibinfo{booktitle}{Proceedings of the 2016 ACM on
  Multimedia Conference}}, MM '16, \bibinfo{pages}{1127--1135},
  \doiprefix\url{10.1145/2964284.2964312} (\bibinfo{publisher}{ACM},
  \bibinfo{address}{New York, NY, USA}, \bibinfo{year}{2016}).

\bibitem{AlbertKG17}
\bibinfo{author}{Albert, A.}, \bibinfo{author}{Kaur, J.} \&
  \bibinfo{author}{Gonz{\'{a}}lez, M.~C.}
\newblock \bibinfo{title}{Using convolutional networks and satellite imagery to
  identify patterns in urban environments at a large scale}.
\newblock In \emph{\bibinfo{booktitle}{Proceedings of the 23rd {ACM} {SIGKDD}
  International Conference on Knowledge Discovery and Data Mining, Halifax, NS,
  Canada, August 13 - 17, 2017}}, \bibinfo{pages}{1357--1366},
  \doiprefix\url{10.1145/3097983.3098070} (\bibinfo{year}{2017}).

\bibitem{stopdl}
\bibinfo{author}{Rudin, C.}
\newblock \bibinfo{journal}{\bibinfo{title}{Stop explaining black box machine
  learning models for high stakes decisions and use interpretable models
  instead}}.
\newblock {\emph{\JournalTitle{Nature Machine Intelligence}}}
  \textbf{\bibinfo{volume}{1}}, \bibinfo{pages}{206--215},
  \doiprefix\url{10.1038/s42256-019-0048-x} (\bibinfo{year}{2019}).

\bibitem{huang2016activity}
\bibinfo{author}{Huang, Q.} \& \bibinfo{author}{Wong, D.~W.}
\newblock \bibinfo{journal}{\bibinfo{title}{Activity patterns, socioeconomic
  status and urban spatial structure: what can social media data tell us?}}
\newblock {\emph{\JournalTitle{International Journal of Geographical
  Information Science}}} \textbf{\bibinfo{volume}{30}},
  \bibinfo{pages}{1873--1898} (\bibinfo{year}{2016}).

\bibitem{adebayo2018sanity}
\bibinfo{author}{Adebayo, J.} \emph{et~al.}
\newblock \bibinfo{title}{Sanity checks for saliency maps}.
\newblock In \emph{\bibinfo{booktitle}{Advances in Neural Information
  Processing Systems}}, \bibinfo{pages}{9505--9515} (\bibinfo{year}{2018}).

\end{thebibliography}

\nocite{*} 

\section*{Acknowledgements}
 This work was supported by the SoSweet ANR project (ANR-15-CE38-0011), the MOTIf Stic-AmSud project (18-STIC-07), and the ACADEMICS project financed by IDEX LYON. 

\section*{Author contributions statement}
All authors designed the research. JLA built the combined dataset and implemented the analysis of the results. All authors wrote the final manuscript. 

\newpage

\setcounter{table}{0}
\renewcommand{\thetable}{S\arabic{table}}%
\setcounter{figure}{0}
\renewcommand{\thefigure}{S\arabic{figure}}%
\setcounter{section}{0}
\renewcommand{\thesection}{S\arabic{section}}%

\hspace{-.2in}{\huge\textbf{Supplementary Information}}\\ \\

\section{Datasets}

\begin{figure}[htbp]
\centerline{\includegraphics[width=\columnwidth]{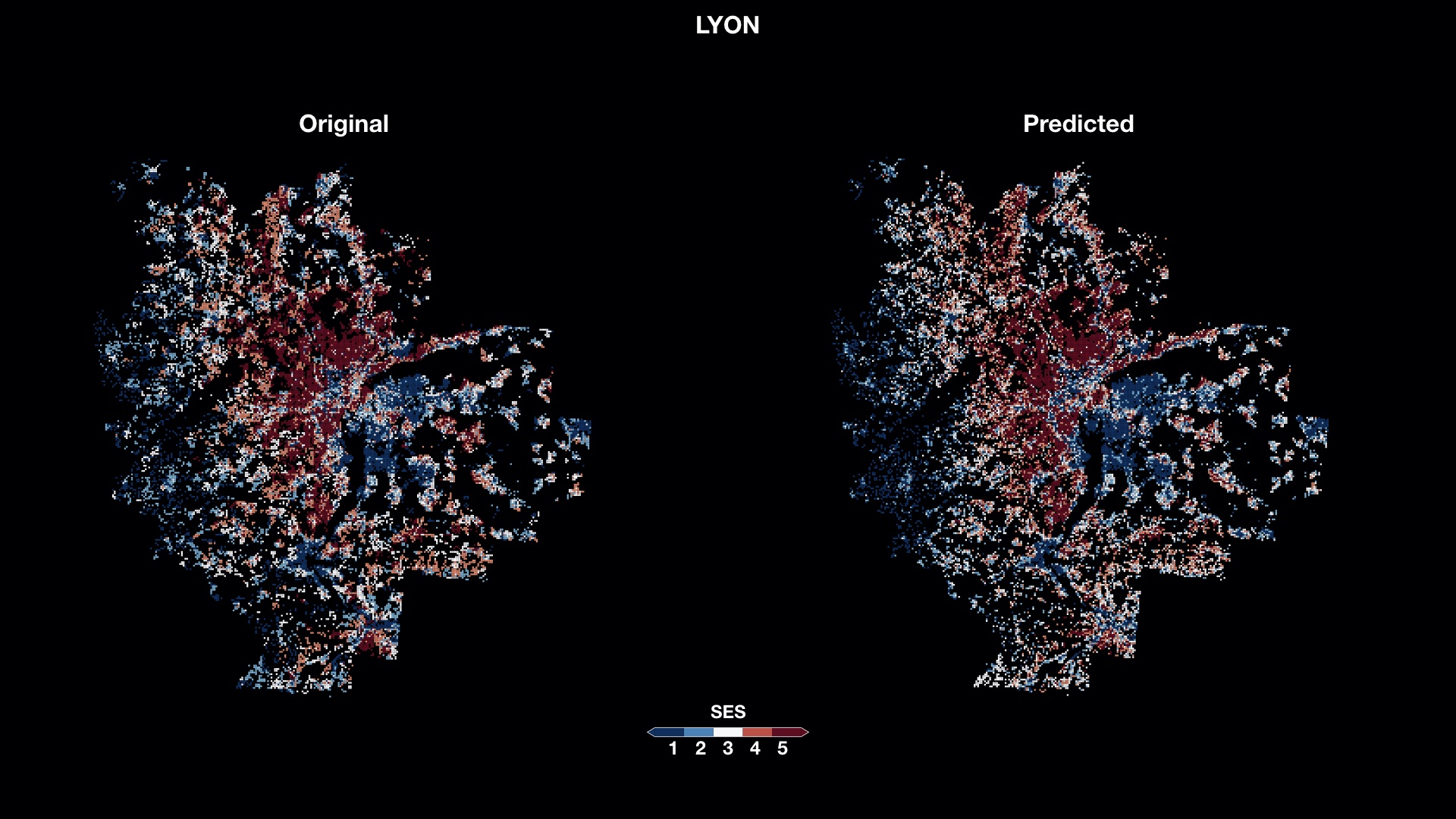}}
\caption{Maps of observed and predicted average income for Lyon.}
\label{fig:dist_lyon}
\end{figure}

\begin{figure}[htbp]
\centerline{\includegraphics[width=\columnwidth]{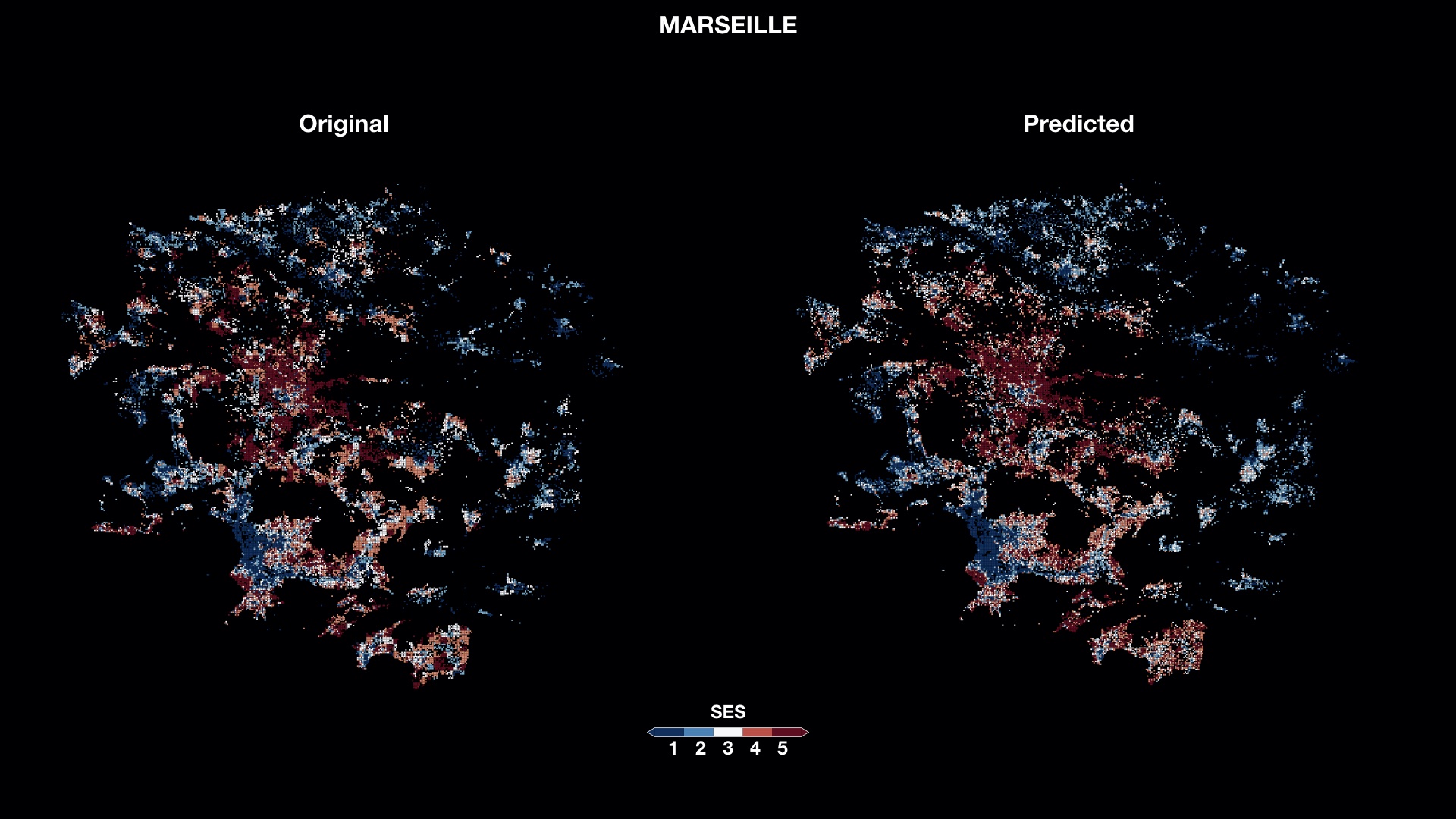}}
\caption{Maps of observed and predicted average income for Marseille. }
\label{fig:dist_marseille}
\end{figure}

\begin{figure}[htbp]
\centerline{\includegraphics[width=\columnwidth]{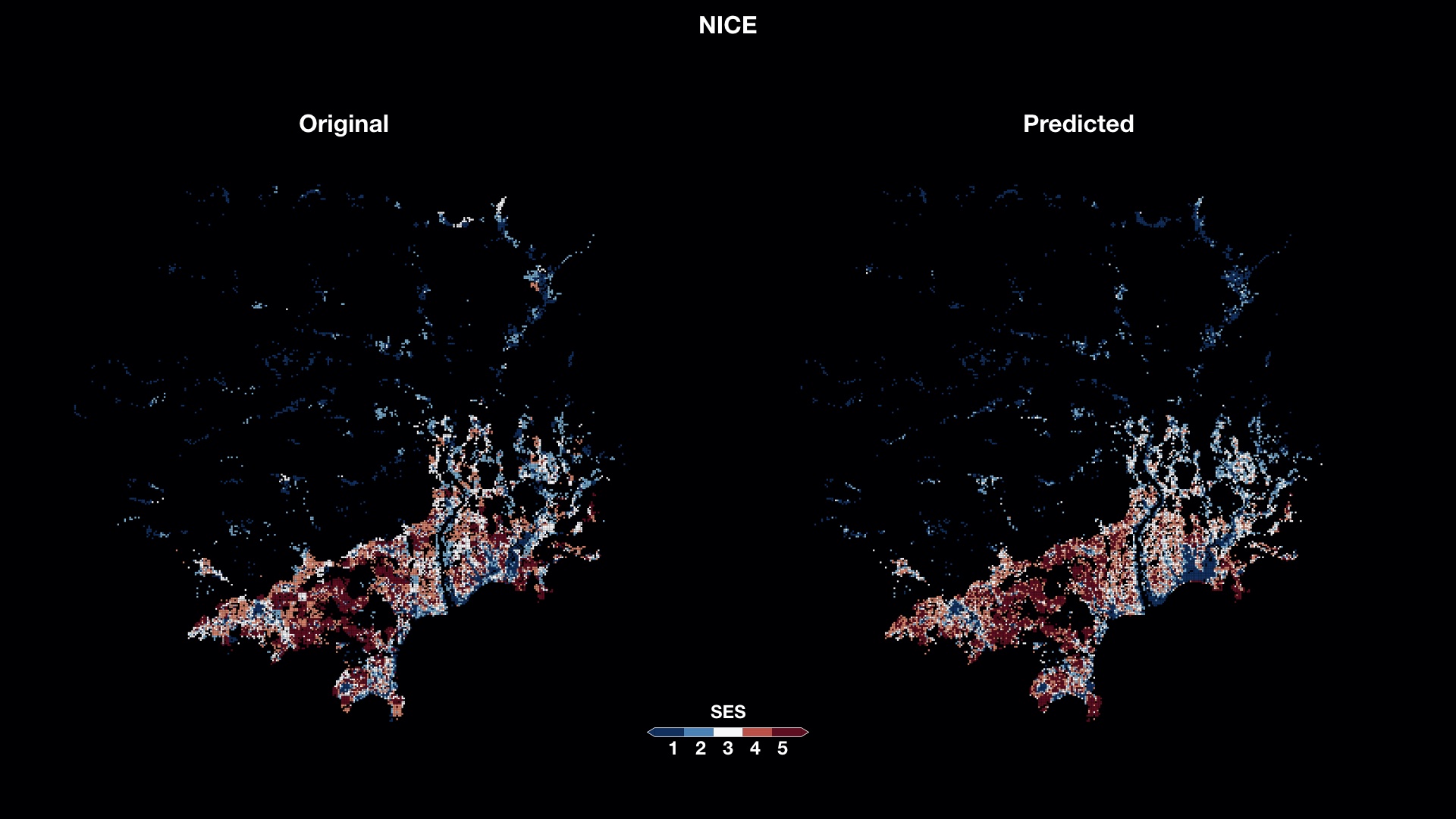}}
\caption{Maps of observed and predicted average income for Nice. }
\label{fig:dist_nice}
\end{figure}

\begin{figure}[htbp]
\centerline{\includegraphics[width=\columnwidth]{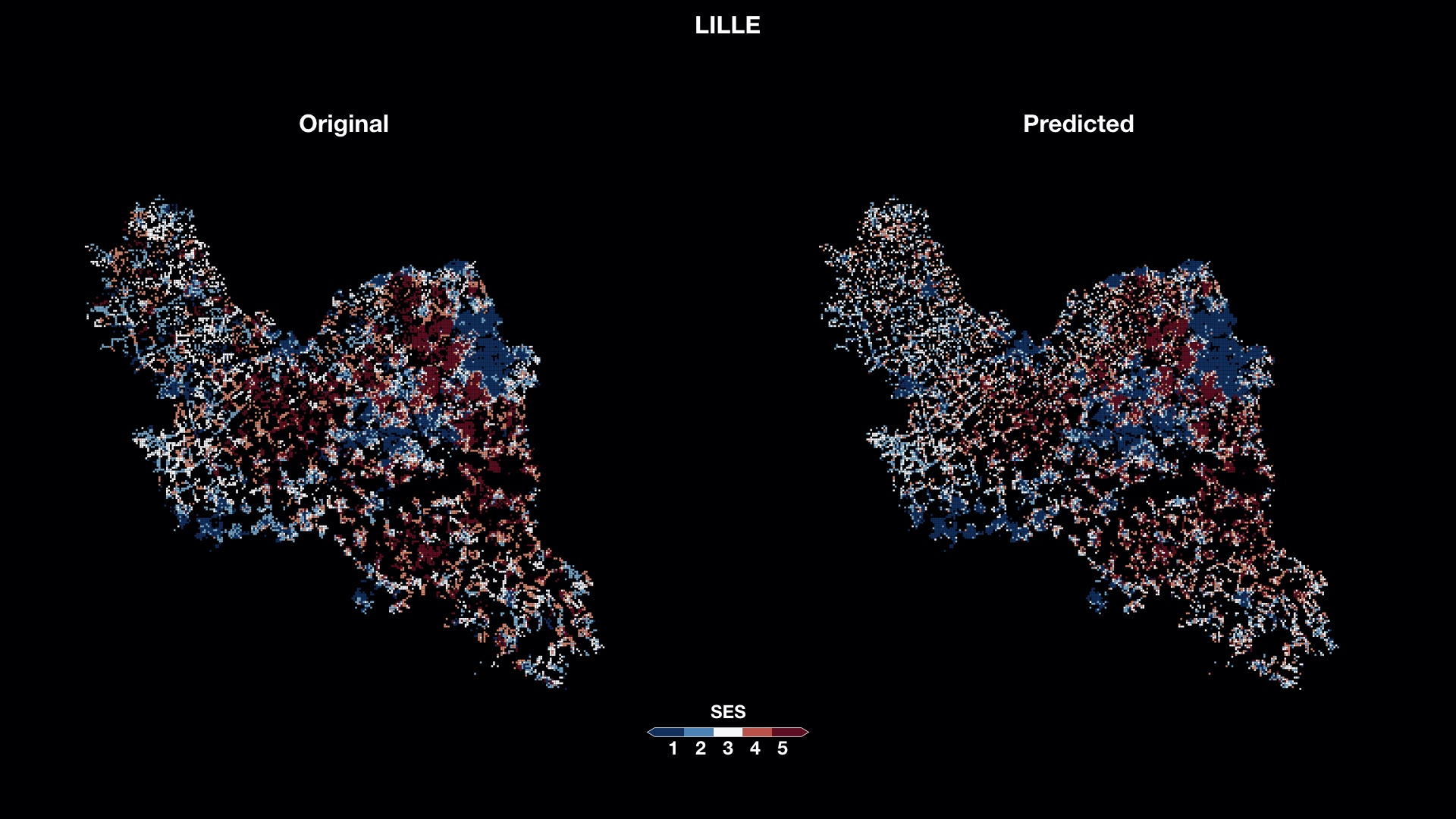}}
\caption{Maps of observed and predicted average income for Lille. }
\label{fig:dist_lille}
\end{figure}

\textbf{Socioeconomic Data.} We obtained detailed sociodemographic information from the French National Institute of Statistics and Economic Studies (INSEE) 2019 gridded dataset~\cite{insee}. This data corpus describes the winsorised average household income, estimated from the 2015 tax return in France, for each 4 hectare (200m x 200m) square patch across the whole French territory (Figure 1b in main text). These socioeconomic patches are also referred to as census cells. Contrary to previous versions of this dataset, the winsorisation is done at the department level, which enables us to get an in-depth view of socioeconomic disparities for the whole country. For each city, income values were partitioned into one of five ($n_{SES}=5$) socioeconomic classes defined by the five quantiles of the city-wise income distribution so that classes 1 and 5 correspond respectively to the bottom and top 20\% of earners in each city. For our study, we singled out 5 major cities located within the French metropolitan territory, namely: \textbf{Paris}, \textbf{Lyon}, \textbf{Marseille}, \textbf{Nice} and \textbf{Lille}. Geographical boundaries for each of these cities/urban areas were obtained from the ensuing Land Use Data. The corresponding spatial distribution of income may be observed in Figures 3a (in the main text), ~\ref{fig:dist_lyon}a,~\ref{fig:dist_marseille}a,~\ref{fig:dist_nice}a, ~\ref{fig:dist_lille}a. where each colored pixel corresponds to a single socioeconomic patch.

\textbf{Aerial Imagery Data.} We obtained high resolution orthophotography taken between 2013 and 2016 of the complete French metropolitan territory (20cm/pixel) from the National Geographical Information Institute (IGN). This dataset is provided as a series of georeferenced 5km x 5km tiles at the department (administrative delimitation) level as shown in Figure 1a. In order to extract the aerial imagery corresponding to each socioeconomic square patch (aerial tiles in what follows), we derived the following procedure: First, for each tile $T$, we identified all the socioeconomic patches contained exclusively within $T$ and we extracted them. Then, we identified all the socioeconomic cells overlapping multiple aerial tiles. These tiles are subsequently merged and the corresponding patches extracted. The results of this procedure can be seen in Figure 1 (left, main text) where we show the image associated to a given socioeconomic patch associated with a particular socioeconomic label. Contrary to previous works~\cite{esra,gebru2017,jean_neal}, we chose not to rely on Google Maps/Streets as a source of data for two basic reasons. First, API crawled data lacks georeferencing which hinders any attempt of overlaying activation maps with land use data. Second, these datasets have recently become harder to access~\cite{maps_price} at the scale needed for conducting this type of work, leading us to eventually discard them. So far, our dataset consists of a series of 200m x 200m aerial tiles fully covering five French cities and each associated to one of five socioeconomic classes.

\textbf{Land Use Data.} We gathered land use information from the 2012 European Union Urban Atlas dataset. This dataset provides high-resolution maps of roughly 700 urban areas of more than 100,000 inhabitants in EU28 and EFTA countries. In doing so, it yields consistent land cover maps encoded via detailed polygons organised in commonly used ESRI shapefiles and covering 27 standardised land use classes (see Figure 1c in main text). This number was later reduced to 19 for this study as infrequent and similar classes were respectively discarded and merged (see explicit list in Table~\ref{table:ua_classes}). Further information regarding the procedure by which this dataset was generated can be obtained from the European Environment Agency~\cite{ua2012}. The land cover information is hence overlaid on top of the image of every socioeconomic patch, yielding results shown in Figure 4c in main text.

Note that all the collected datasets can be openly accessed and emanate from a single 4-year time window hence reducing temporal misalignment between them to a minimum.

\section{Abbreviations table}
We provide here an overview of the abbreviations used in plots containing urban classes descriptors:

\begin{center}
    \begin{tabular}{| l | p{9cm} |}
    \hline
    \textbf{ID}& Full Denomination \\ \hline
    \textbf{1}& Agricultural areas\\ \hline
    \textbf{2}& Green UA (Urban Areas) \\ \hline
    \textbf{3}& Natural Areas\\ \hline
    \textbf{4}& Industrial/Commercial Areas\\ \hline
    \textbf{5}& Sports/Leisure Facilities\\ \hline
    \textbf{6}& Construction/Dump Sites\\ \hline
    \textbf{7}& No Use (Land without Current Use) \\ \hline
    \textbf{8}& Roads \\ \hline
    \textbf{9}& Railway\\ \hline
    \textbf{10}& Motorways\\ \hline
    \textbf{11}& Port Areas\\ \hline
    \textbf{12}& Water\\ \hline
    \textbf{13}& Open Spaces with little or no vegetation\\ \hline
    \textbf{14}& Isolated Structures\\ \hline
    \textbf{15}& Very Low Density Discontinuous Urban Fabric (s.l. < 10\%)\\ \hline
    \textbf{16}& Low Density Discontinuous Urban Fabric (s.l. 10\% - 30\%)\\ \hline
    \textbf{17}& Medium Density Discontinuous Urban Fabric (s.l. 30\% - 50\%)\\ \hline
    \textbf{18}& Dense Discontinuous Urban Fabric (s.l. 50\% - 80\%)\\ \hline
    \textbf{19}& Continuous Urban Fabric (s.l. > 80\%)\\ \hline
    \end{tabular}
    \label{table:ua_classes}
    \captionof{table}{Land cover classes (i.e urban classes) used in this study.}
\end{center}

\section{Socioeconomic Predictions for other cities}
\label{dist_other}

\newpage

\section{Guided Grad-CAM (GG-CAM)}
Grad-CAM generates coarse, discriminative regions for each of the socioeconomical classes the models were trained to predict. It is calculated as the rectified linear  of the weighted sum of the feature maps $A^k$ from the last convolutional layer (\verb conv2d_65  in the original EfficientNetB0 model) of the CNN: 
$$ L^c_{\text{
Grad-CAM}} = \text{ReLU} \Big(\sum_{k}\alpha_k^cA^k\Big)$$
This  weighted combination of forward activation maps is based on the weights $\alpha_k^c$, defined by :
$$\alpha_k^c =  \frac{1}{Z}\sum_{i}\sum_{j}\frac{\partial y^c}{\partial A_{i,j}^k}$$
where $Z$ is a normalization constant, $y^c$ is the prediction score for SES class $c$ and $A_{i,j}^k$ is the $ij^{\text{th}}$ element of $A^k$. Each $\alpha_k^c$ represents a partial linearization of the CNN downstream from $A^k$, and captures the importance of the feature map $k$ for a SES class $c$. Once computed, activation maps are $l^1$-normalized. Guided BackPropagation, on the other hand, is a method that captures non class-discriminative details of visual components that are of some importance to the network by suppressing the flow of gradients through neurons wherein either of input or incoming gradients were negative. It is computed as :
$$ R^l = f^{l'}\text{ ReLU }\Big(\frac{\partial f^{out}}{\partial f^{l'}}\Big)\frac{\partial f^{out}}{\partial f^{l'}}$$
with $R_l$ guided backpropagation product of the $l$-th layer, $f_l$ and $f^{\text{out}}$ feature maps respectively  of the $l-$th and last convolutional layer and $f^{l'}=\text{ ReLU }(f^l)$.
Guided GradCAM are then generated via the Hadamard product of GradCAM and the guided backpropagation saliency map. Contrary to previous works, the outcome of this operation isn't normalized any further so as to enable comparison between samples from the same city. Furthermore, to take into account recent work by Adebayo et al.~\cite{adebayo2018sanity}, no absolute value is applied to pixel activation values. Rather, values are truncated to zero and only considered when not lying in the top or bottom 1\% of the activation distribution.

\newpage

\section{Univariate Activation Ratios and Empirical Probabilities for other cities}

\begin{figure}[ht!]
\centerline{\includegraphics[width=.85\columnwidth]{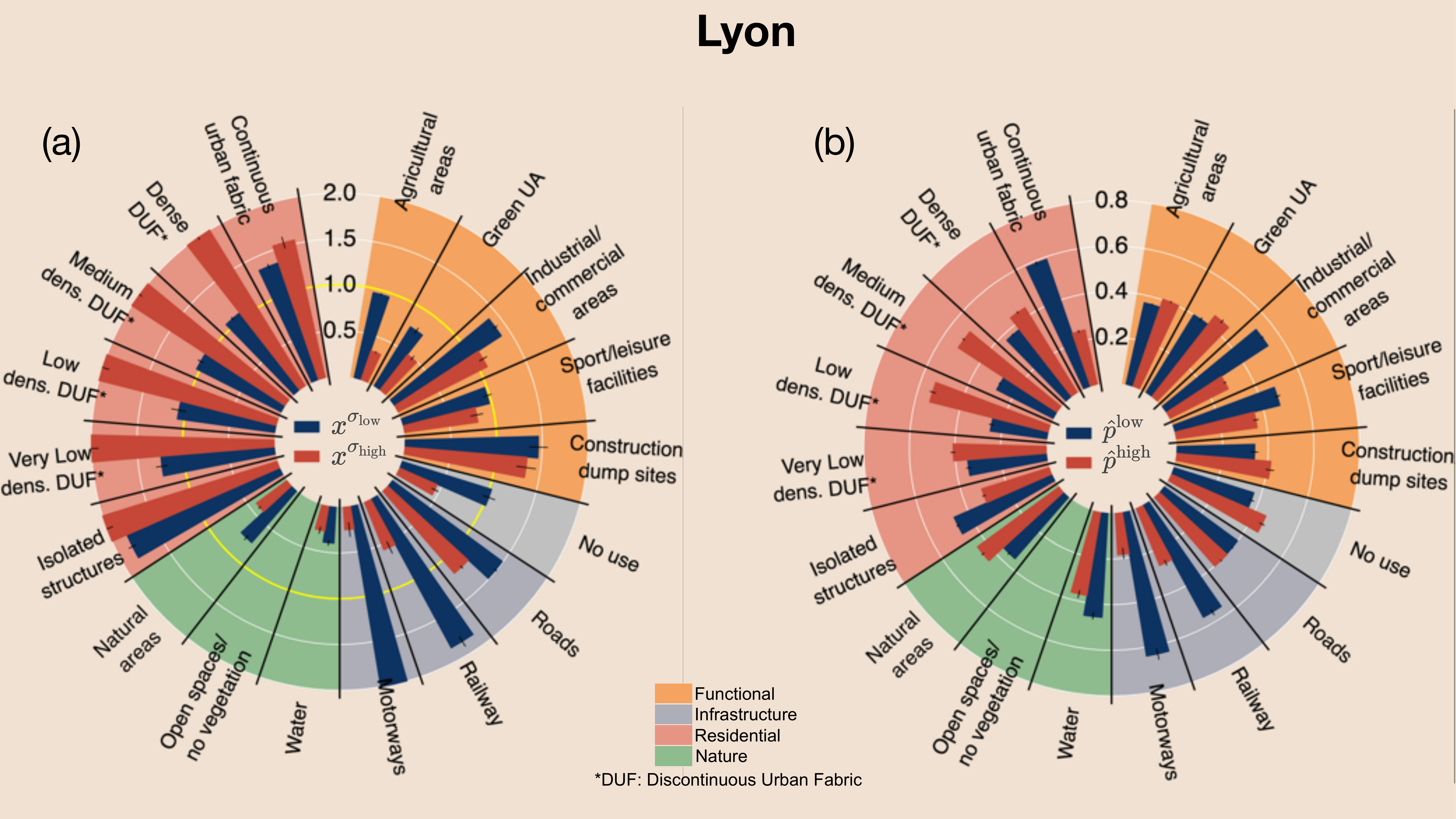}}
\caption{Correlations between urban topology and socioeconomic status in the city of Lyon: (a) Mean model activation rate per urban class with bootstrapped 95\% confidence interval for samples predicted as respectively as low SES (blue) or high SES (red) by the model (b) Estimated probability of an urban polygon belonging to the bottom or top quintile of the income distribution with bootstrapped  95\% confidence interval.}
\end{figure}

\begin{figure}[ht!]
\centerline{\includegraphics[width=.85\columnwidth]{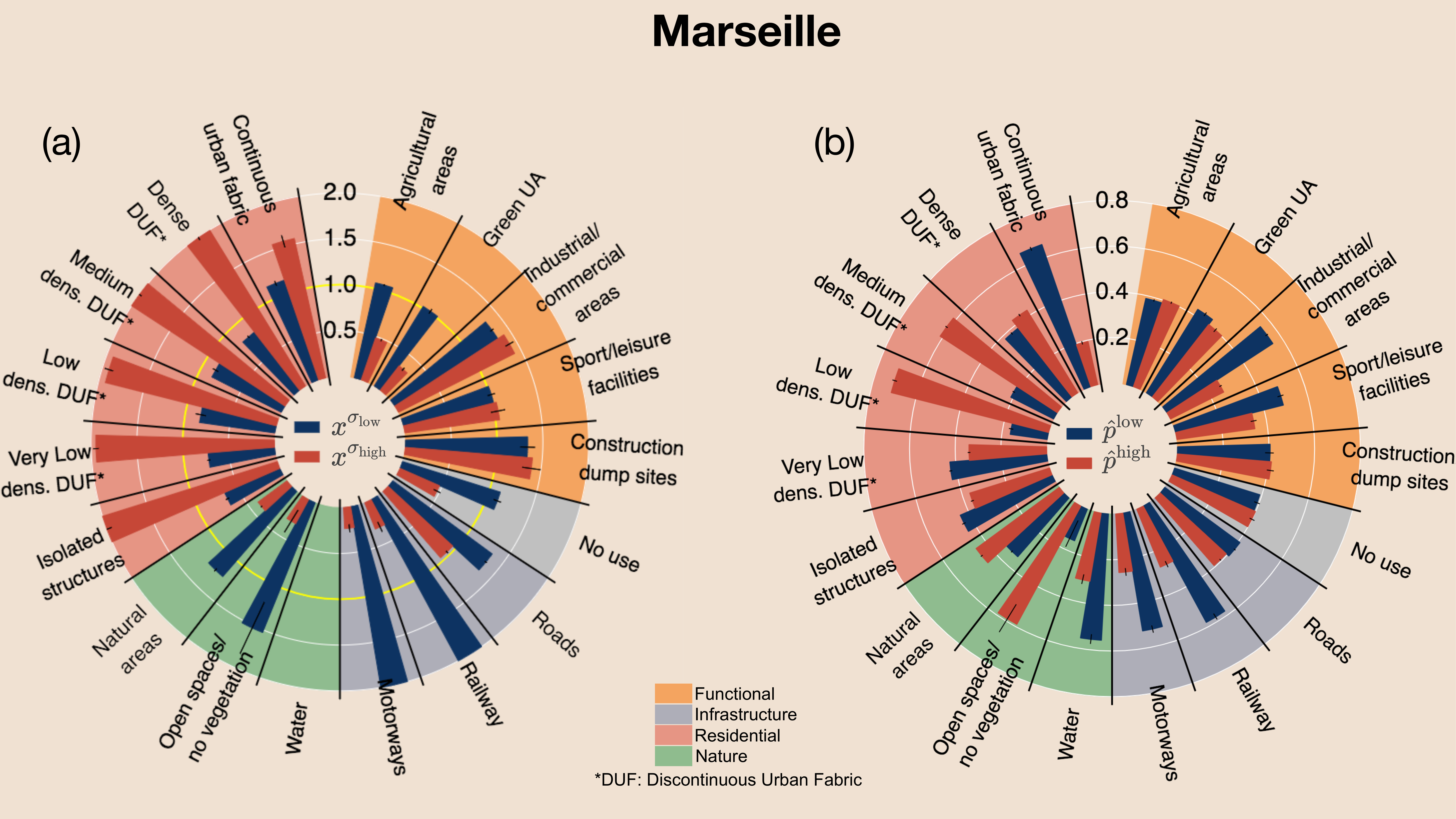}}
\caption{Correlations between urban topology and socioeconomic status in the city of Marseille: (a) Mean model activation rate per urban class with bootstrapped 95\%  confidence interval for samples predicted as respectively as low SES (blue) or high SES (red) by the model (b) Estimated probability of an urban polygon belonging to the bottom or top quintile of the income distribution with bootstrapped  95\% confidence interval.}
\end{figure}

\begin{figure}[ht!]
\centerline{\includegraphics[width=.85\columnwidth]{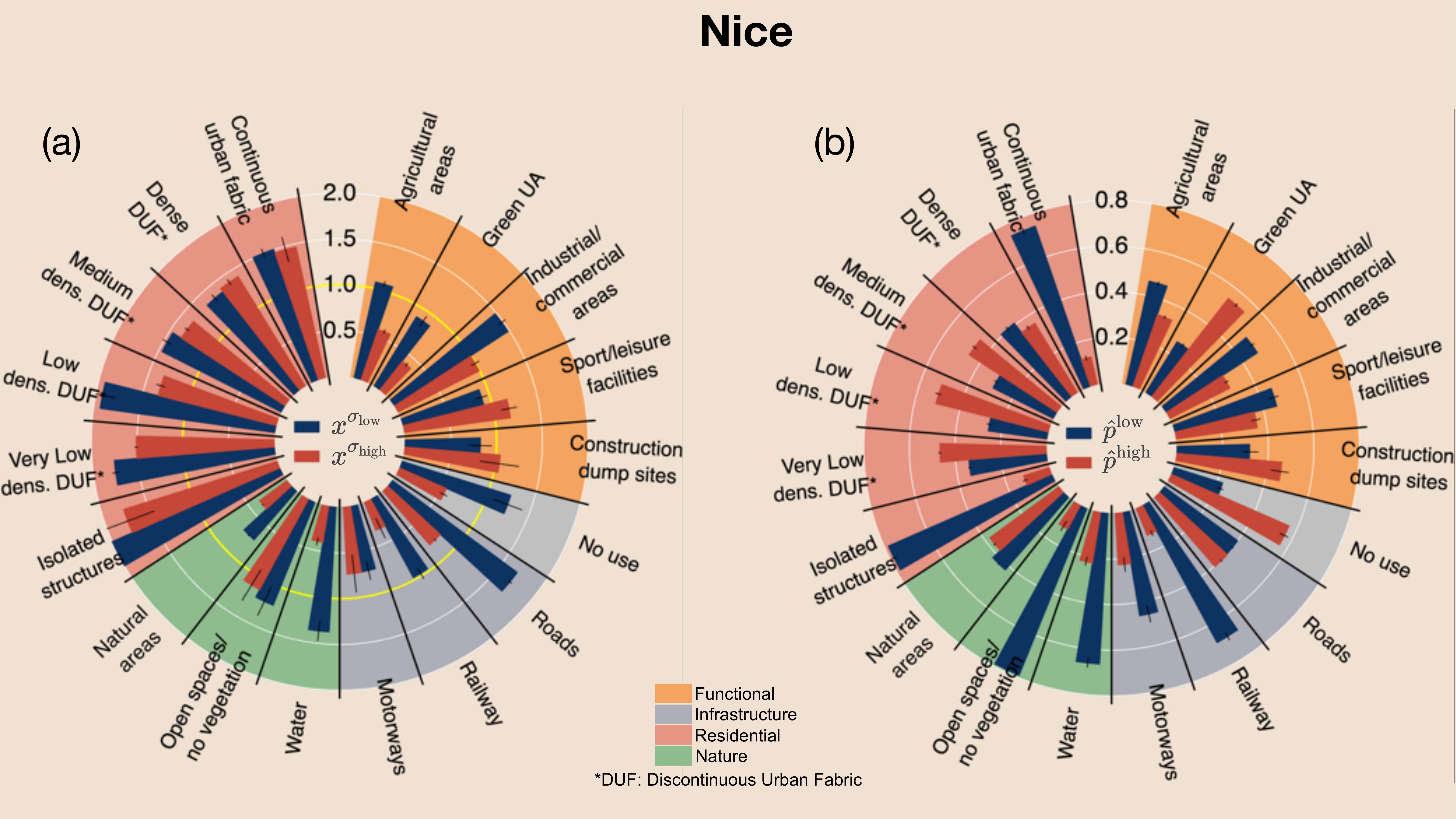}}
\caption{Correlations between urban topology and socioeconomic status in the city of Nice: (a) Mean model activation rate per urban class with bootstrapped 95\% confidence interval for samples predicted as respectively as low SES (blue) or high SES (red) by the model (b) Estimated probability of an urban polygon belonging to the bottom or top quintile of the income distribution with bootstrapped  95\% confidence interval.}
\end{figure}

\begin{figure}[ht!]
\centerline{\includegraphics[width=.85\columnwidth]{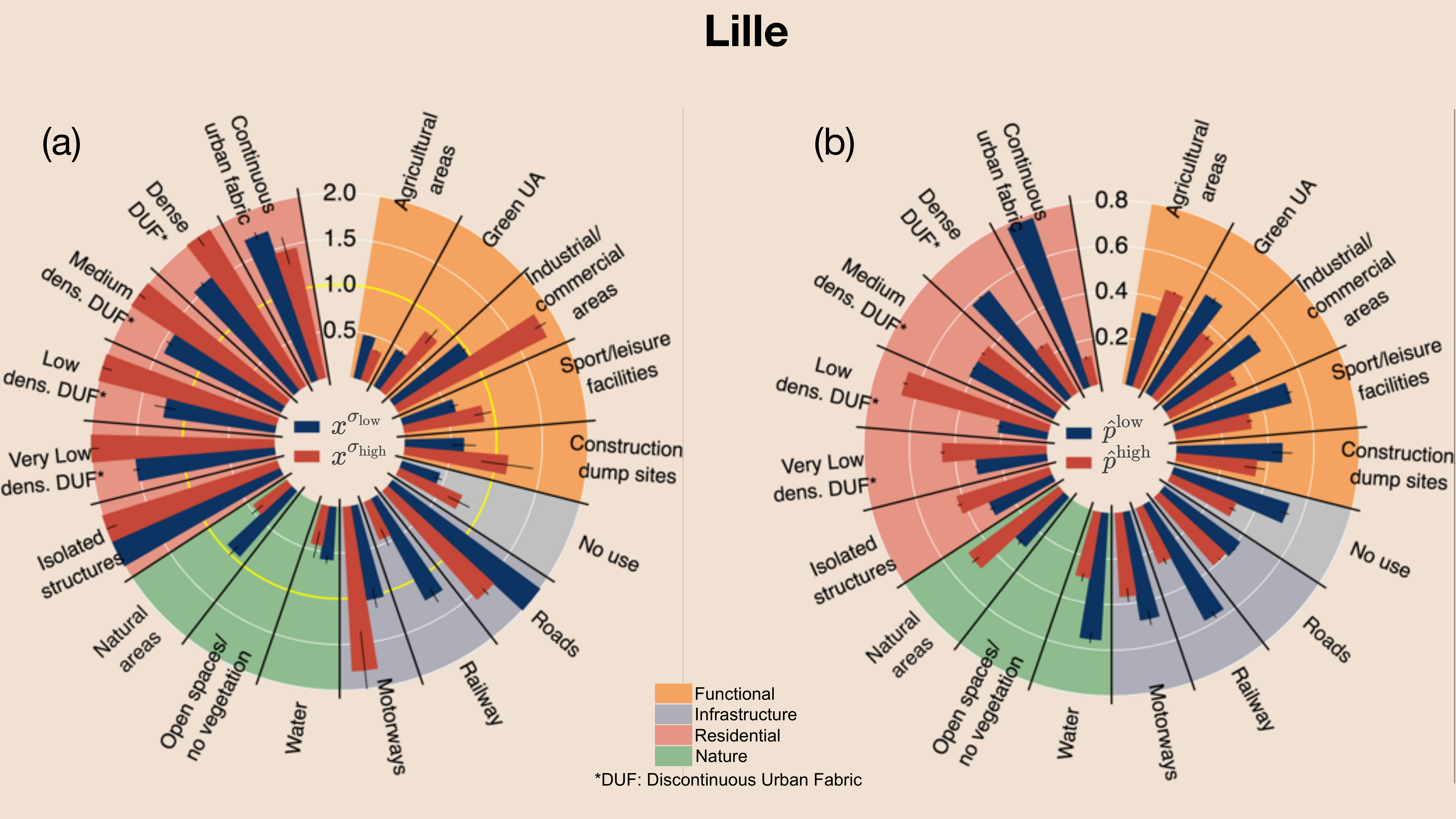}}
\caption{Correlations between urban topology and socioeconomic status in the city of Lille: (a) Mean model activation rate per urban class with bootstrapped 95\% confidence interval for samples predicted as respectively as low SES (blue) or high SES (red) by the model (b) Estimated probability of an urban polygon belonging to the bottom or top quintile of the income distribution with bootstrapped  95\% confidence interval.}
\end{figure}

\begin{figure}[htbp]
\centerline{\includegraphics[height=0.9\textheight]{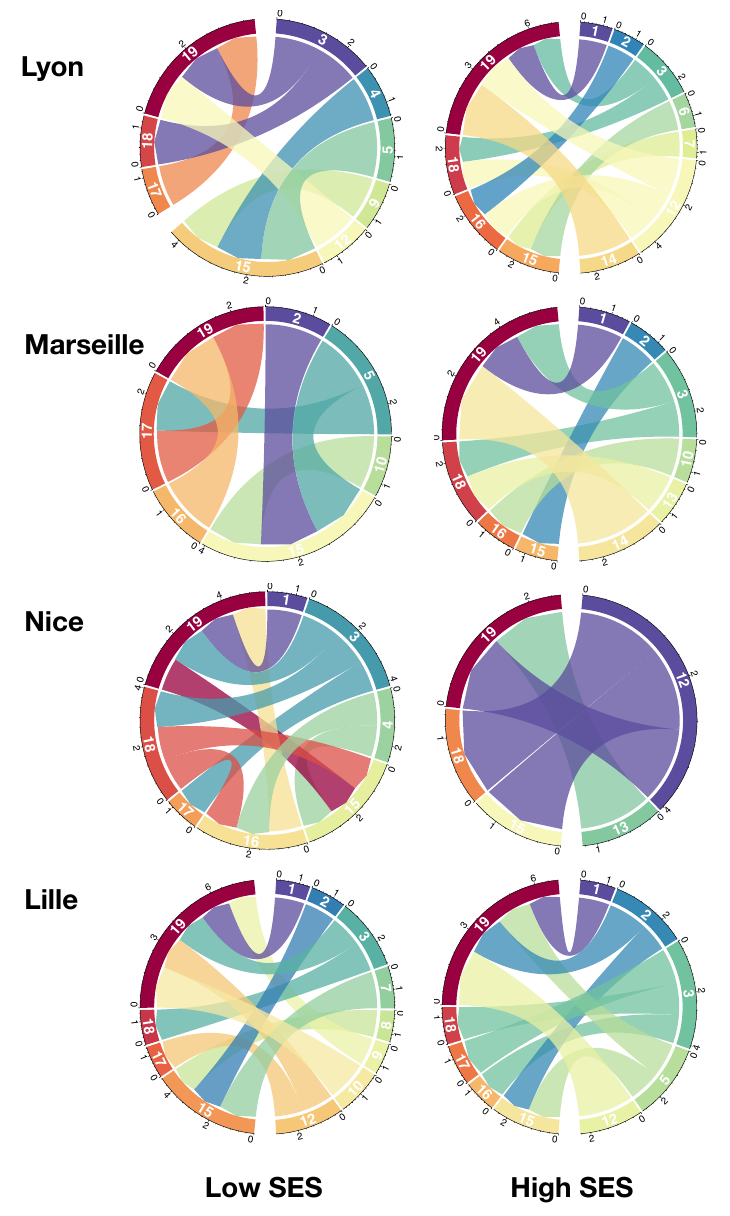}}
\caption{ Chord diagrams of coactivations estimated for both low (left) and high (right) SES in the remaining set of cities. Whenever urban class $i$ appears more activated for a given SES in the presence of urban class $j$ than when appearing by itself or any other random urban class, it is represented in the diagram as $j\rightarrow i$. Each circular segment represents an urban class $i$ connected by chords with width equal to the coactivation value. Urban classes are labeled based on their ID (Table~\ref{table:ua_classes}). Colors are assigned in order of appearance for better visualization based on a common colormap and are not exclusive to a given class. For simplicity, we illustrate only pairs of classes with coactivations  values greater than 1.2 (at least 20\% improvement over the univariate case). Among these only the top ten ranked by coactivations are shown.}
\end{figure}

\begin{figure}[htbp]
\centerline{\includegraphics[height=0.9\textheight]{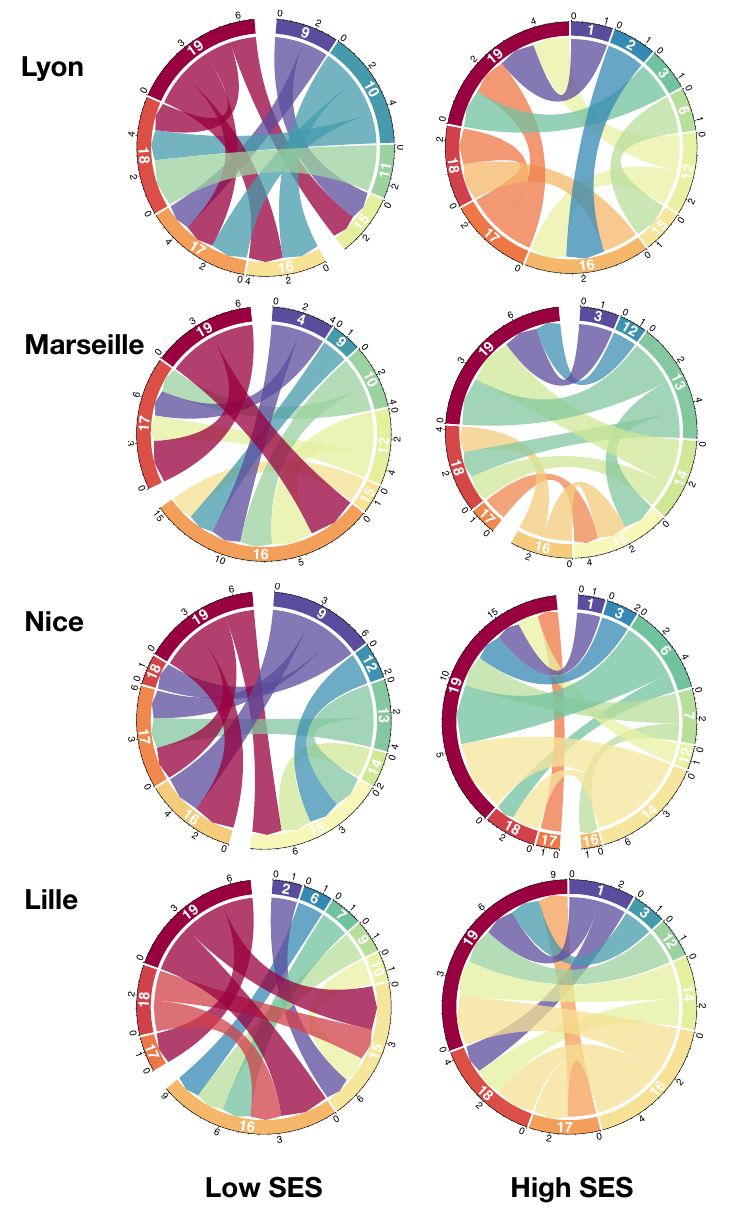}}
\caption{Chord diagrams of coappearance gains estimated for both low (left) and high (right) SES in the remaining set of cities. Whenever urban class $i$ is more likely to be of a given SES in the presence of urban class $j$ than when appearing by itself or any other random urban class, it is represented in the diagram as $j\rightarrow i$. Each circular segment represents an urban class $i$ connected by chords with width equal to the coappearance gain. Urban classes are labeled based on their ID (Table~\ref{table:ua_classes}). Colors are assigned in order of appearance for better visualization based on a common colormap and are not exclusive to a given class. For simplicity, we illustrate only pairs of classes with coappearance gains values greater than 1.2 (at least 20\% improvement over the univariate case). Among these only the top ten ranked by coappearance gains are shown.}
\end{figure}

\end{document}